\documentclass[twocolumn,english]{revtex4}
\usepackage[T1]{fontenc}
\usepackage[latin9]{inputenc}
\usepackage{color}
\usepackage{amsmath}
\usepackage{amssymb}
\usepackage{graphicx}
\usepackage{esint}

\makeatletter
\@ifundefined{textcolor}{}
{%
 \definecolor{BLACK}{gray}{0}
 \definecolor{WHITE}{gray}{1}
 \definecolor{RED}{rgb}{1,0,0}
 \definecolor{GREEN}{rgb}{0,1,0}
 \definecolor{BLUE}{rgb}{0,0,1}
 \definecolor{CYAN}{cmyk}{1,0,0,0}
 \definecolor{MAGENTA}{cmyk}{0,1,0,0}
 \definecolor{YELLOW}{cmyk}{0,0,1,0}
 }

\newcommand{\mfr}[2]{\left|\tiny{\begin{smallmatrix}#1\\ #2 \end{smallmatrix}}\right\rangle}

\usepackage{babel}

\usepackage{babel}

\usepackage{babel}

\makeatother

\usepackage{babel}
\begin{document}

\title{Frustrated Ising model on the Cairo pentagonal lattice}

\author{M. Rojas, Onofre Rojas and S. M. de Souza }

\affiliation{Departamento de Ciencias Exatas, Universidade Federal de Lavras,
C.P. 3037, 37200-000 Lavras,  Minas Gerais, Brazil },
\begin{abstract}
Through the direct decoration transformation approach, we obtain a
general solution for the pentagonal Ising model, showing its equivalence
to the isotropic free-fermion eight-vertex model. We study
the ground-state phase diagram, in which one ferromagnetic (FM) state,
one ferrimagnetic (FIM) state, and one frustrated state are found.
Using the exact solution of the pentagonal Ising model, we discuss
the finite-temperature phase diagrams and find a phase transition
between the FIM state and the disordered state as well as a phase
transition between the disordered state and the FM state. We also
discuss some additional remarkable properties of the model, such as
the magnetization, entropy, and specific heat, at finite temperature
and at its low-temperature asymptotic limit. Because of the influence
of the second-order phase transition between the frustrated and ferromagnetic
phases, we obtain surprisingly low values of the entropy and the specific
heat until the critical temperature is reached. 
\end{abstract}

\pacs{05.10.-a; 05.50.+q; 75.10.Hk; 64.60.De;}

\keywords{Ising model; Exactly solvable model; Geometric frustration.}

\maketitle

\section{Introduction}

Over the past six decades, much effort has been devoted to determining
the critical behavior of statistical properties of lattice models,
which would allow a deeper understanding of order-disorder phenomena
in magnetic solids. Following Onsager's pioneering exact solution for
the square lattice Ising model \cite{on}, exact solutions were also
 obtained for other regular two-dimensional lattice structures \cite{gr}.
In particular, exact results have been attained for the triangular,
honeycomb, kagome, and bathroom-tile lattices \cite{ho}, \cite{sy},
\cite{uti}, as well as for two-dimensional models, such as the Union Jack
(centered square) \cite{va} and the square kagome \cite{sun} lattices.

Geometrical frustration is mainly based on the triangle and tetrahedron
structures, but it was also found in the Ising model on a pentagonal
Penrose lattice proposed by Waldor \textit{et al.}\cite{wal}  and solved exactly using the transfer matrix approach.

More recently Urumov \cite{uru}  considered the Ising model on
the Cairo pentagonal lattice using the decoration transformation \cite{fisher}.
This model has been mapped onto the Union Jack lattice \cite{va}
and its critical temperature and spontaneous magnetization properties
have been discussed. This model is interesting from a mathematical
point of view. A few years ago, real materials with a Cairo  pentagonal
lattice structure were found; for example, the $\mathrm{Fe^{3+}}$
lattice in $\mathrm{Bi_{2}Fe_{4}O_{9}}$ (described as a pentagonal
Heisenberg model) was discussed by Ressourche \textit{et al}. \cite{cairo}.
This material shows magnetic frustration. Also, theoretical calculations
of the phonon structure of  antiferromagnetic $\mathrm{Bi_{2}Fe_{4}O_{9}}$
(space group  \textit{Pbnm} No. 55, $T\approx240\mathrm {K}$) were studied using
 lattice dynamics and these results were confirmed experimentally
by polarized Raman spectroscopy from 10 to 300 K \cite{Iliev}. More
recently, some additional experimental studies were performed \cite{Liu,K Jin}.  Ralko \cite{cairo-hubb} also discussed the hard-core extended
boson Hubbard model on the Cairo pentagonal lattice, using the numerical
quantum Monte Carlo study of  stochastic series expansion and cluster
mean-field theory.

The purpose of this paper is to present a general exact solution of
the pentagonal Ising model and as a special case  we obtain the Urumov
solution using a standard decoration transformation approach \cite{fisher}.
Furthermore, we present a more simplified solution through the direct
decoration transformation \cite{ono}  instead of the standard one \cite{fisher}.
The generalized version of the latter \cite{phys-A-09,strecka-pla}
is widely used to solve some two-dimensional decorated Ising\cite{strecka-mixd}
and Ising-Heisenberg \cite{our-4-6-latt,strecka-triang,loh} models.

This paper is organized as follows. In Sec. II  we consider the
detailed description of the Ising model on the Cairo pentagonal lattice.
In Sec. III we discuss its phase diagram at zero temperature. Section
IV is devoted to the pentagonal Ising model mapping, using the direct
decoration transformation \cite{ono} for the isotropic free-fermion
vertex model \cite{Fan-wu}, presenting the most relevant results and
discussion. In Sec. V  we obtain the finite-temperature phase
diagrams, critical temperature, magnetization, entropy, and specific
heat.  Section VI summarizes our discussion.

\section{Ising model on the Cairo pentagonal lattice}

\begin{figure}[h]
\begin{centering}
\includegraphics[scale=0.4,angle=-90]{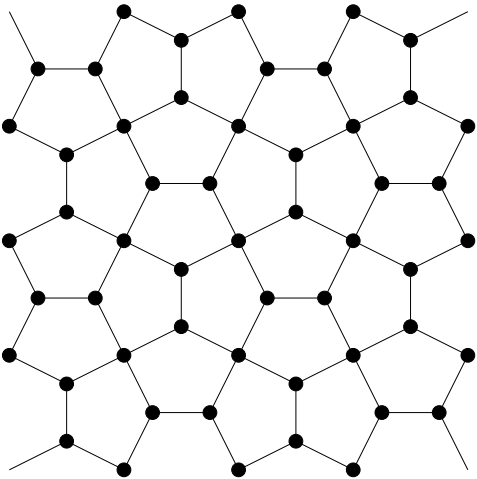} 
\par\end{centering}

\caption{\label{fig:pent-latt}Schematic representation of the Cairo pentagonal
lattice.}
\end{figure}

The highly anisotropic Heisenberg model on the Cairo pentagonal lattice
considered by Ressourche   \textit{et al}. \cite{cairo}  could be reduced to the Ising
model on a Cairo pentagonal lattice. Therefore, let us consider the
Ising model on a planar lattice where the tiling is achieved with
non regular pentagons; the lattice may be viewed as an assembly of
checkerboard ordering with the elementary cell (see Fig. \ref{fig:transf-pent})
rotated by $\pi/2$ in the neighboring square plaquettes, as shown
in Fig. \ref{fig:pent-latt}   (for more details see Ref. \cite{uru}).

The Hamiltonian of the Cairo pentagonal Ising model,  (represented
schematically in Fig. \ref{fig:pent-latt}), discussed previously  by Urumov \cite{uru},
is expressed by 
\begin{equation}
H=-J_{1}\sum_{\langle i,j \rangle}s_{i}s_{j}-J\sum_{\langle k,l \rangle}s_{k}\tau_{l},\label{eq:hmt-1}
\end{equation}
where the first summation is the contribution of the interaction between
the nearest neighbor with spin $s_{i}$ ($s_{i}$ interacting with
coordination number 3) and $J_{1}$ corresponds to the interaction between
$s_{i}$ and $s_{j}$. While the second summation is the contribution
of the nearest-neighbor interaction $J$ between spin $s_{k}$ and
spin $\tau_{l}$ ($\tau_{l}$'s interacting with coordination number
4), conveniently we assume $s_{i}=\pm1$ and $\tau_{l}=\pm1$.

\section{Phase diagram at zero temperature}

In this section we discuss the phase diagram at zero temperature
of the Hamiltonian given in Eq. \eqref{eq:hmt-1}.

In order to discuss the phase diagram at zero temperature, we define
the magnetization $M$ for the pentagonal lattice that is used throughout
the paper, given by 
\begin{equation}
M=\frac{M_{0}+2M_{1}}{3},\label{eq:T-mag}
\end{equation}
with $M_{0}=\langle\tau_{1}\rangle$ and $M_{1}=\langle s_{1}\rangle$.

The energy per plaquette of three ground states  that appear for the pentagonal
lattice Ising model is expressed in terms of an elementary cell (see
Fig. \ref{fig:transf-pent}). It is worth  highlighting that the
elementary cell should not be confused with
the unit cell of the pentagonal lattice.

(i) The ferromagnetic (FM) state or saturated state has a total magnetization
$M=1$ and ground-state energy per plaquette $E=-J_{1}-4J$. Thus
the FM state can be represented as 
\begin{equation}
|\mathrm{FM}\rangle=\mfr{+_{+}+}{+^{+}+}.
\end{equation}
This state is limited by $J>0$ for $J_{1}>0$ and\textcolor{green}{{}
}$J_{1}>-J$ for $J_{1}<0$, as displayed in Fig. \ref{fig:Ph-dgm}.

(ii) The ferrimagnetic state (FIM) has a total magnetization $M=1/3$ and
ground-state energy per plaquette $E=-J_{1}+4J$, which corresponds
to the configuration displayed in Fig. \ref{fig:Ph-dgm}. Analogous
to the previous case, we describe the state by 
\begin{equation}
|\mathrm{FIM}\rangle=\mfr{+_{-}+}{+^{-}+}.
\end{equation}
This state is limited by $J<0$ for $J_{1}>0$ and $J_{1}>J$ for
$J_{1}<0$, as illustrated in Fig. \ref{fig:Ph-dgm}.

(iii) The frustrated state (FRU) is given as a combination of states
$\mfr{\sigma_{\sigma}\sigma}{\sigma^{-\sigma}\sigma}$ with its rotated
elementary cell and spin inversion on the elementary cell with ground-state energy per plaquette $E=J_{1}-2|J|$, which can be expressed
by 
\begin{equation}
|\mathrm{FRU}\rangle=\text{combinations\,\ of}\;\{\mfr{+_{+}+}{+^{-}-},\mfr{-_{-}-}{-^{+}+}\}.\label{eq:fru-st}
\end{equation}
This state is limited by $J_{1}\leqslant-|J|$ (see Fig. \ref{fig:Ph-dgm}).

\begin{figure}
\begin{centering}
\includegraphics[scale=0.22]{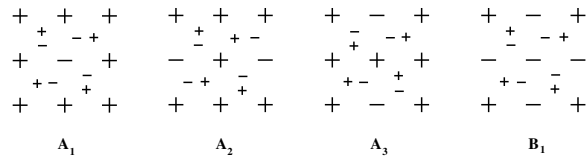} 
\par\end{centering}

\caption{\label{fig:sublatt}Unit cells $A_{i}$ ($i=1,2,3$) with magnetization
$M=1/6$ and unit cell $B_{1}$ with $M=-1/6$.}
\end{figure}

\begin{figure}
\begin{centering}
\includegraphics[clip]{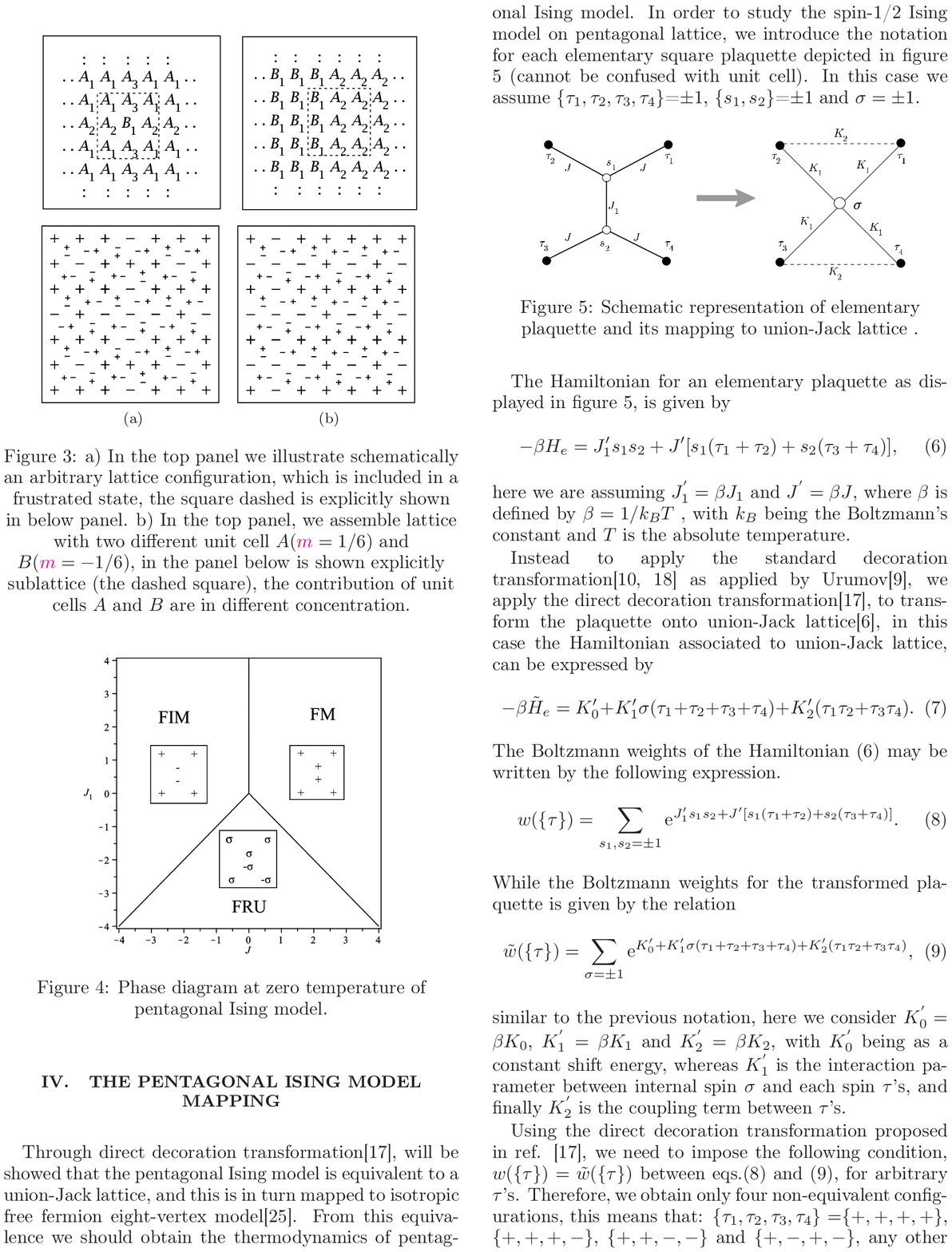} 
\par\end{centering}

\caption{\label{fig:latticeAB}(a) In the top (left) panel, we illustrate schematically
an arbitrary lattice configuration, which is included in a frustrated
state; the dashed square  is shown explicitly  in the panel below. ( b)
In the top (right) panel, we assemble a lattice with two different unit
cells $A$($m=1/6)$ and $B$($m=-1/6$). In the panel below, the
sublattice (the dashed square) is explicitly shown; the contributions
of unit cells $A$ and $B$ are in different concentrations. }
\end{figure}

We define  $m$ as the magnetization for each frustrated-state configuration in the
range between $m=-1/6$ and $1/6$ ( we  denote by $M$ the
average of total magnetization). Combining the state of the elementary
cell displayed in Eq. \eqref{eq:fru-st} and its rotation in $\pi/2$
of the elementary cell, it is possible to generate the geometrically
frustrated state. In particular, when we combine  half states
$\mfr{+_{+}+}{+^{-}-}$ and the remaining states with $\mfr{-_{-}-}{-^{+}+}$,
we obtain an antiferromagnetic  state, with null total magnetization.
Other intermediate states with magnetization $0\leqslant m\leqslant1/6$
also could be obtained by combining the elementary cell state with
different relative amounts of the state given by Eq. \eqref{eq:fru-st}.
More specifically, the unit cell magnetization could be classified
as displayed in Fig. \ref{fig:sublatt}, i.e, $A_{i}$ ($i=1,2,3$)
and $B_{1},$ with magnetization $m=1/6$ and $-1/6$, respectively.
Certainly, this is not the only way to classify the unit cell by its
magnetization;   any other classifications of the unit cell
lead us to the same kind of lattice configuration. In Fig. \ref{fig:latticeAB}
we show two particular situations of such a lattice configuration, formed
by the unit cells of type $A_{i}$ and $B$. In the bulk limit we
 have a lattice with total magnetization $1/6$.  In Fig. \ref{fig:latticeAB}(b)  the lattice is composed of unit cells $A_{2}$ and $B_{1}$ with a different concentration,
generating a different total magnetization of the lattice. Thus we
obtain the total magnetization of each particular configuration. However,
the average magnetization of all these configurations with equal energy
will be null $M=0$ of the frustrated state in the interval $-1/6<m<1/6$.

\begin{figure}[h]
\begin{centering}
\includegraphics[scale=0.3]{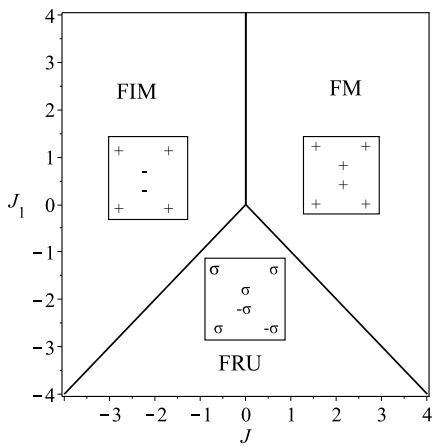} 
\par\end{centering}

\caption{\label{fig:Ph-dgm}Phase diagram at zero temperature of the pentagonal
Ising model. }
\end{figure}

\section{The pentagonal Ising model mapping}

\begin{figure}[h]
\begin{centering}
\includegraphics[scale=0.5]{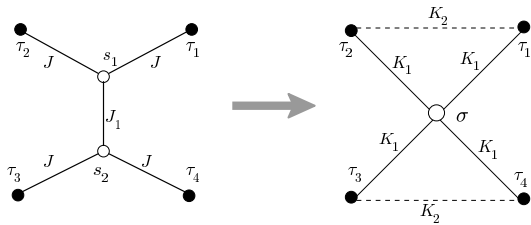} 
\par\end{centering}

\caption{\label{fig:transf-pent}Schematic representation of the elementary plaquette
and its mapping to the Union Jack lattice .}
\end{figure}

Through the direct decoration transformation \cite{ono}, it is shown that
the pentagonal Ising model is equivalent to a Union Jack lattice,
which is in turn mapped onto the isotropic free-fermion eight-vertex
model \cite{ch,baxter-free-ferm}. From this equivalence we can obtain the thermodynamics
of the pentagonal Ising model. In order to study the spin-1/2 Ising model
on a pentagonal lattice, we introduce the notation for each elementary
square plaquette depicted in Fig. \ref{fig:transf-pent} (this should not
be confused with the unit cell). In this case we assume $\{\tau_{1},\tau_{2},\tau_{3},\tau_{4}\}$=$\pm1$,
$\{s_{1},s_{2}\}$=$\pm1$ and $\sigma=\pm1$.

The Hamiltonian for an elementary plaquette as displayed in Fig.
\ref{fig:transf-pent}  is given by 
\begin{equation}
-\beta H_{e}=J'_{1}s_{1}s_{2}+J'[s_{1}(\tau_{1}+\tau_{2})+s_{2}(\tau_{3}+\tau_{4})],\label{pa}
\end{equation}
where we are assuming $J_{1}^{'}=\beta J_{1}$ and $J^{'}=\beta J$ and $\beta$ is defined by $\beta=1/k_{B}T$, with $k_{B}$ 
the Boltzmann constant and $T$  the absolute temperature.

Instead of applying the standard decoration transformation \cite{fisher,phys-A-09}
as applied by Urumov \cite{uru}, we apply the direct decoration transformation \cite{ono}
to transform the plaquette into the Union Jack lattice \cite{va}.
In this case the Hamiltonian associated with Union Jack lattice can
be expressed by 
\begin{equation}
-\beta\tilde{H}_{e}=K'_{0}+K'_{1}\sigma(\tau_{1}+\tau_{2}+\tau_{3}+\tau_{4})+K'_{2}(\tau_{1}\tau_{2}+\tau_{3}\tau_{4}).\label{H2_eff}
\end{equation}
The Boltzmann weights of the Hamiltonian \eqref{pa} may be written
 
\begin{equation}
{\displaystyle w(\{\tau\})=\sum_{s_{1},s_{2}=\pm1}\mathrm{e}^{J'_{1}s_{1}s_{2}+J'[s_{1}(\tau_{1}+\tau_{2})+s_{2}(\tau_{3}+\tau_{4})]}},\label{eq:w1}
\end{equation}
whereas the Boltzmann weights for the transformed plaquette is given
by the relation 
\begin{equation}
\tilde{w}(\{\tau\})=\sum_{\sigma=\pm1}\mathrm{e}^{K'_{0}+K'_{1}\sigma(\tau_{1}+\tau_{2}+\tau_{3}+\tau_{4})+K'_{2}(\tau_{1}\tau_{2}+\tau_{3}\tau_{4})}.\label{eq:w2}
\end{equation}
Similar to the previous notation, here we consider $K_{0}^{'}=\beta K_{0}$,
$K_{1}^{'}=\beta K_{1}$ and $K_{2}^{'}=\beta K_{2}$, where $K_{0}^{'}$
is taken as a constant shift energy,  $K_{1}^{'}$ is the
interaction parameter between the internal spin $\sigma$ and each spin
$\tau$, and  $K_{2}^{'}$ is the coupling term between spins $\tau$.

Using the direct decoration transformation proposed in Ref. \cite{ono},
we need to impose the  condition, $w(\{\tau\})=\tilde{w}(\{\tau\})$
between Eqs. \eqref{eq:w1} and \eqref{eq:w2} for arbitrary $\tau$.
Therefore, we obtain only four non equivalent configurations $\{\tau_{1},\tau_{2},\tau_{3},\tau_{4}\}=$$\{+,+,+,+\}$,
$\{+,+,+,-\}$, $\{+,+,-,-\}$, and $\{+,-,+,-\}$; any other permutation
or spin inversion falls onto one of these configurations. Thus the
Boltzmann weight for each configuration is given by 
\begin{align}
\xi_{1}=w(+,+,+,+)= & 2\mathrm{e}^{K'_{0}+2K'_{2}}\cosh(4K_{1}^{\prime}),\label{eq:rot-w1}\\
\xi_{2}=w(+,-,+,-)= & 2\mathrm{e}^{K'_{0}-2K'_{2}},\\
\xi_{3}=w(+,+,-,-)= & 2\mathrm{e}^{K'_{0}+2K'_{2}},\\
\xi_{5}=w(+,+,+,-)= & 2\mathrm{e}^{K'_{0}}\cosh(2K'_{1}),\label{eq:rot-w5}
\end{align}
\textcolor{black}{where $\xi_{2}=\xi_{4}$ and $\xi_{5}=\xi_{6}=\xi_{7}=\xi_{8}$. }

\begin{figure}
\includegraphics[scale=0.33]{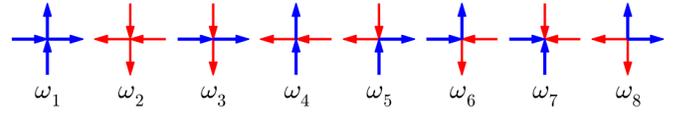}\caption{\label{fig:8-vertex}(Color online) Eight-vertex model diagrams}
\end{figure}

The above equations satisfy the isotropic free-fermion condition \cite{Fan-wu}\textcolor{black}{{}
$w_{1}w_{2}+w_{3}w_{4}=w_{5}w_{6}+w_{7}w_{8}$, following the eight-vertex
model with Boltzmann weights $\omega_{1},...,\omega_{8}$ ($\omega$
should not be confused with $w$) displayed in} Fig. \ref{fig:8-vertex}.
Hence the free-fermion condition may be rewritten in terms of $\xi$
as  
\begin{equation}
2\xi_{5}^{2}=(\xi_{1}+\xi_{3})\xi_{2}.\label{Exact-cond}
\end{equation}

Therefore, the Boltzmann factor of an effective Union Jack lattice can
be expressed in terms of the pentagonal Ising model coupling parameters
\begin{align}
\xi_{1}= & ru^{-4}+2r^{-1}+ru^{4},\\
\xi_{2}= & 2\left(r+r^{-1}\right),\\
\xi_{3}= & 2r+r^{-1}u^{-4}+r^{-1}u^{4},\\
\xi_{5}= & \left(r+r^{-1}\right)\left(u^{2}+u^{-2}\right).
\end{align}
For simplicity we used the  notation $r=\mathrm{e}^{J'_{1}}$
and $u=\mathrm{e}^{J'}$.

Due to the step by step decoration transformation performed by Urumov \cite{uru}\textcolor{black}{{}
(see Fig. 2),} $J_{1}$ was restricted only to $J_{1}>0$ or $r>1$
(ferromagnetic);  otherwise, if we consider $J_{1}<0$,
we get a imaginary parameter in the intermediate transformation proposed
by Urumov. However, using the direct decoration transformation, we
do not have such a  restriction (Fig. \ref{fig:transf-pent}), but only
$r>0$, which means that $J_{1}$ exchange coupling could be ferromagnetic
or antiferromagnetic.

Hence the pentagonal Ising model is completely equivalent to the
Ising model on the Union Jack lattice \cite{va} with the isotropic
nearest-neighbor interactions defined by $K_{1}$ and non crossing
diagonal interactions between the second nearest neighbor given by
$K_{2}$. In contrast, the Union Jack lattice was mapped onto the
isotropic free-fermion eight-vertex model \cite{Fan-wu} by Choy and
Baxter \cite{ch}. Therefore, we relate the Boltzmann factor given
by Eqs. (\ref{eq:rot-w1})-(\ref{eq:rot-w5}) and the Boltzmann factor
of the Union Jack lattice given by Eq. (4) of Ref. \cite{ch}. These
relations are given by 
\begin{equation}
\omega_{1}=\frac{2\xi_{1}}{\sqrt{\xi_{2}\xi_{3}}},\;\omega_{2}=\frac{2\xi_{2}}{\sqrt{\xi_{2}\xi_{3}}},\;\omega_{3}=2\;\mathrm{and}\;\omega_{5}=\frac{2\xi_{5}}{\sqrt{\xi_{2}\xi_{3}}}.\label{eq:ome}
\end{equation}
The schematic representation of the eight-vertex model is given in
Fig. \ref{fig:8-vertex}.

These Boltzmann weights will be used in the following section to study
the critical temperature and spontaneous magnetization of the Cairo
pentagonal Ising model.

\section{Thermodynamics of pentagonal lattice}

In this section we discuss  thermodynamical properties, such as
the entropy, specific heat, and magnetization, as a function of temperature,
as well as the critical temperature behavior. The thermodynamics of the pentagonal
Ising model can be expressed following the results given by Fan and
Wu \cite{Fan-wu}. The exact result for the free energy of the pentagonal
Ising model is then given by 
\begin{equation}
\beta f=-\frac{1}{4\pi}\int_{0}^{2\pi}\ln\left[A(\phi)+\sqrt{Q(\phi)}\right]\mathrm{d}\phi,\label{eq:mr}
\end{equation}
where 
\begin{align}
A(\phi)= & \tfrac{1}{2}\left(\xi_{1}^{2}+\xi_{2}^{2}+2\xi_{2}\xi_{3}\right)+(\xi_{1}-\xi_{2})\sqrt{\xi_{2}\xi_{3}}\cos(\phi),\label{eq:a1}\\
Q(\phi)= & \left[(\xi_{1}-\xi_{2})\sqrt{\xi_{2}\xi_{3}}\cos(\phi)+\tfrac{1}{2}\left(\xi_{1}+\xi_{2}\right)^{2}\right]^{2}\nonumber \\
 & +\xi_{1}\xi_{2}\left(4\xi_{2}\xi_{3}-\left(\xi_{1}+\xi_{2}\right)^{2}\right).\label{eq:q1}
\end{align}
Once  the free energy is known, we can obtain straightforwardly the critical
temperature, magnetization, entropy, and  specific heat.

\subsection{Critical temperature}

In order to study the spontaneous magnetization, following the result
obtained by Choy and Baxter \cite{ch} and using Eq. (\ref{eq:ome}),
the magnetization $M_{0}=\langle\tau_{1}\rangle$ for spins with coordination
number 4 is described by 
\begin{equation}
M_{0}=\sqrt[8]{1-k^{2}},\label{eq:mag2}
\end{equation}
where 
\begin{equation}
k=\frac{2\xi_{2}\left(\xi_{1}+\xi_{3}\right)}{\xi_{1}^{2}+\xi_{2}^{2}-2\xi_{2}\xi_{3}}.\label{eq:k11}
\end{equation}

Equation (\ref{eq:k11}) is expressed in terms of the pentagonal Ising
model Boltzmann factor. It is important to note that this relation
is valid for arbitrary spins $s_{1}$ and $s_{2}$, as shown in Fig.
\ref{fig:transf-pent}. Using the results obtained in Ref. \cite{ch},
the critical point of the Union Jack lattice is obtained from the
condition $w_{1}-w_{2}=2w_{3}$ for $w_{1}>w_{2}$ or $w_{2}-w_{1}=2w_{3}$
for $w_{2}>w_{1}$;  in terms of $k$, this means that the critical
points occur at $k=1$. Equivalently, using the pentagonal Ising model
Boltzmann factor $\xi$, we have 
\begin{equation}
\left(\xi_{1}-\xi_{2}\right)^{2}=4\xi_{2}\xi_{3}.\label{eq:k22}
\end{equation}
This condition must satisfy the critical point. Rewriting Eq.
(\ref{eq:k22}) in terms of $r$ and $u$, the critical points must
satisfy the  relation 
\begin{align}
r_{c} & =\sqrt{2}u_{c}\sqrt{\frac{2u_{c}^{6}+2u_{c}^{2}+(1-u_{c}^{4})^{2}\sqrt{2}}{\left(u_{c}^{12}-5u_{c}^{8}-5u_{c}^{4}+1\right)}},\label{eq:kk}
\end{align}
where  $r_{c}$ and $u_{c}$  denote  $r$ and $u$ evaluated
at the critical temperature $T_{c}$. The same expression could be obtained from Eq.(8) of the Ref.
\cite{uru}; however, due to the standard decoration transformation \cite{fisher,phys-A-09}  
used by Urumov\cite{uru} for this model, we have to eliminate the
intermediate parameter $Q$. Once  the intermediate parameter
$Q$ is eliminated from Eq. (8) of Ref.  \cite{uru}, it becomes identical to
our Eq. \eqref{eq:kk}  for the case of $J_{1}>0$ (ferromagnetic
coupling), which was previously studied by Urumov \cite{uru}.

\begin{figure}[h]
\begin{centering}
\includegraphics[scale=0.4]{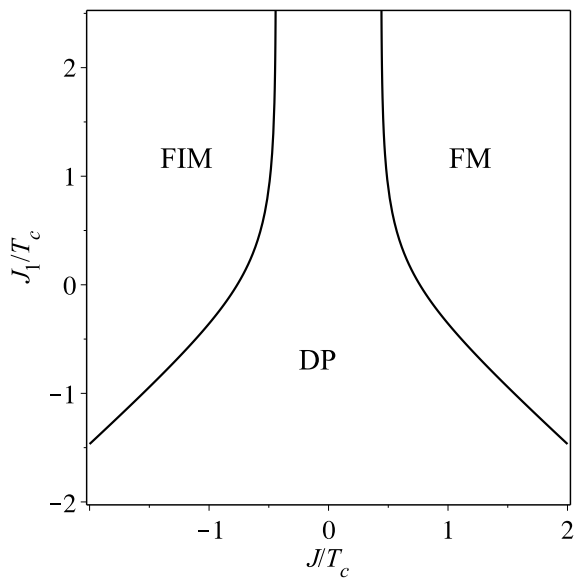} 
\par\end{centering}

\caption{\label{fig:crit-points}Critical points curve fo the pentagonal Ising
model  as a function of the parameters $J/T_{c}$ and $J_{1}/T_{c}$.}
\end{figure}

In Eq. (\ref{eq:kk}) we provide a closed expression for the critical
point of the pentagonal Ising model. The curves where the critical
points occur  are illustrated in Fig. \ref{fig:crit-points}.

The finite-temperature properties of the system arer investigated
by considering the effect of parameters $J$ and $J_{1}$  on the critical
behavior. In Fig. \ref{fig:crit-points} we display the critical
point  regions in units of critical temperature, where the phase transitions
of the FIM region to a disordered  phase (DP) and a DP
to a  FM phases are illustrated. It is important to highlight that Fig.
\ref{fig:crit-points} becomes similar to Fig. \ref{fig:Ph-dgm}
when $T_{c}\rightarrow0$.

\begin{figure}[h]
\begin{centering}
\includegraphics[scale=0.4]{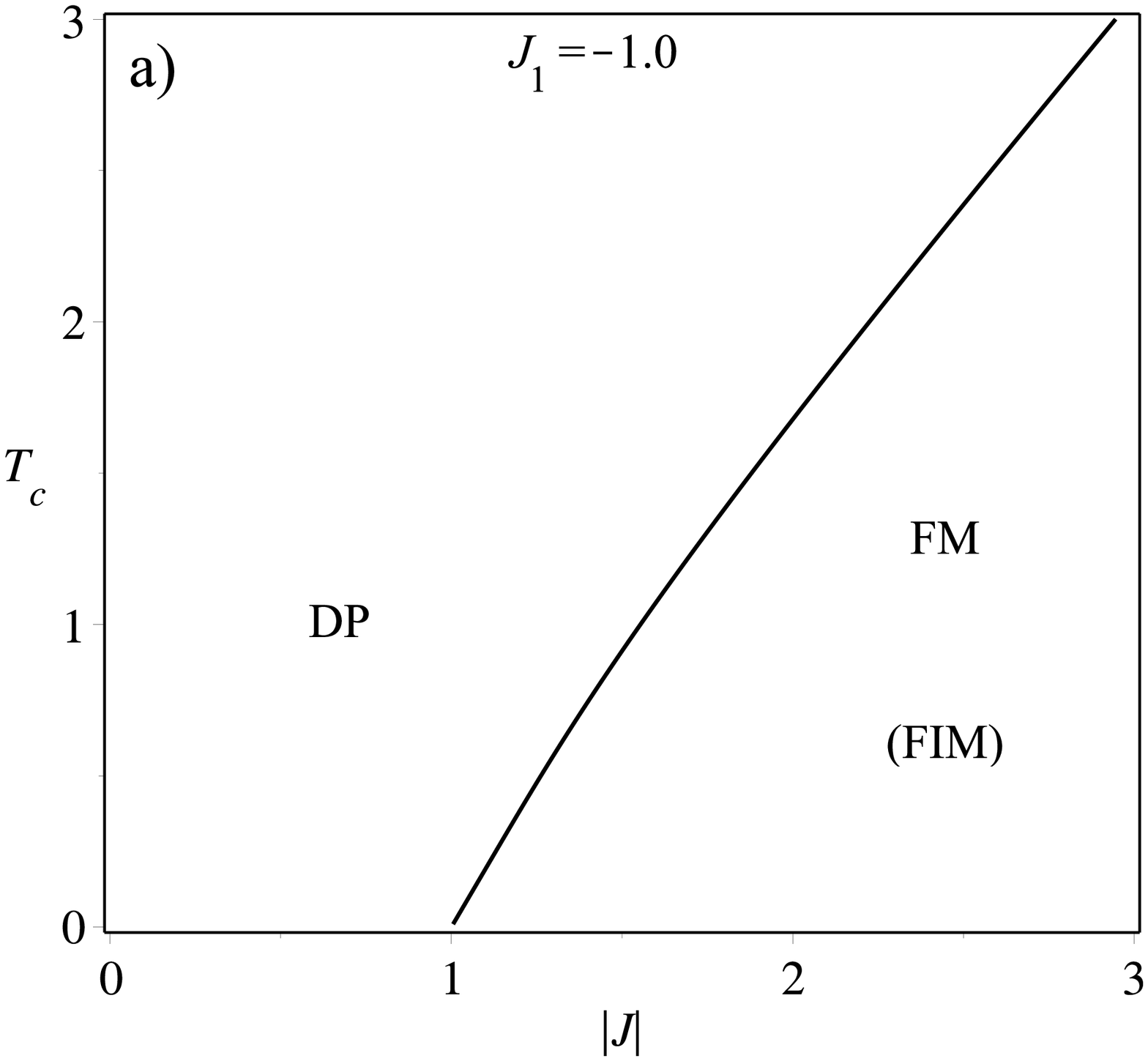} \includegraphics[scale=0.4]{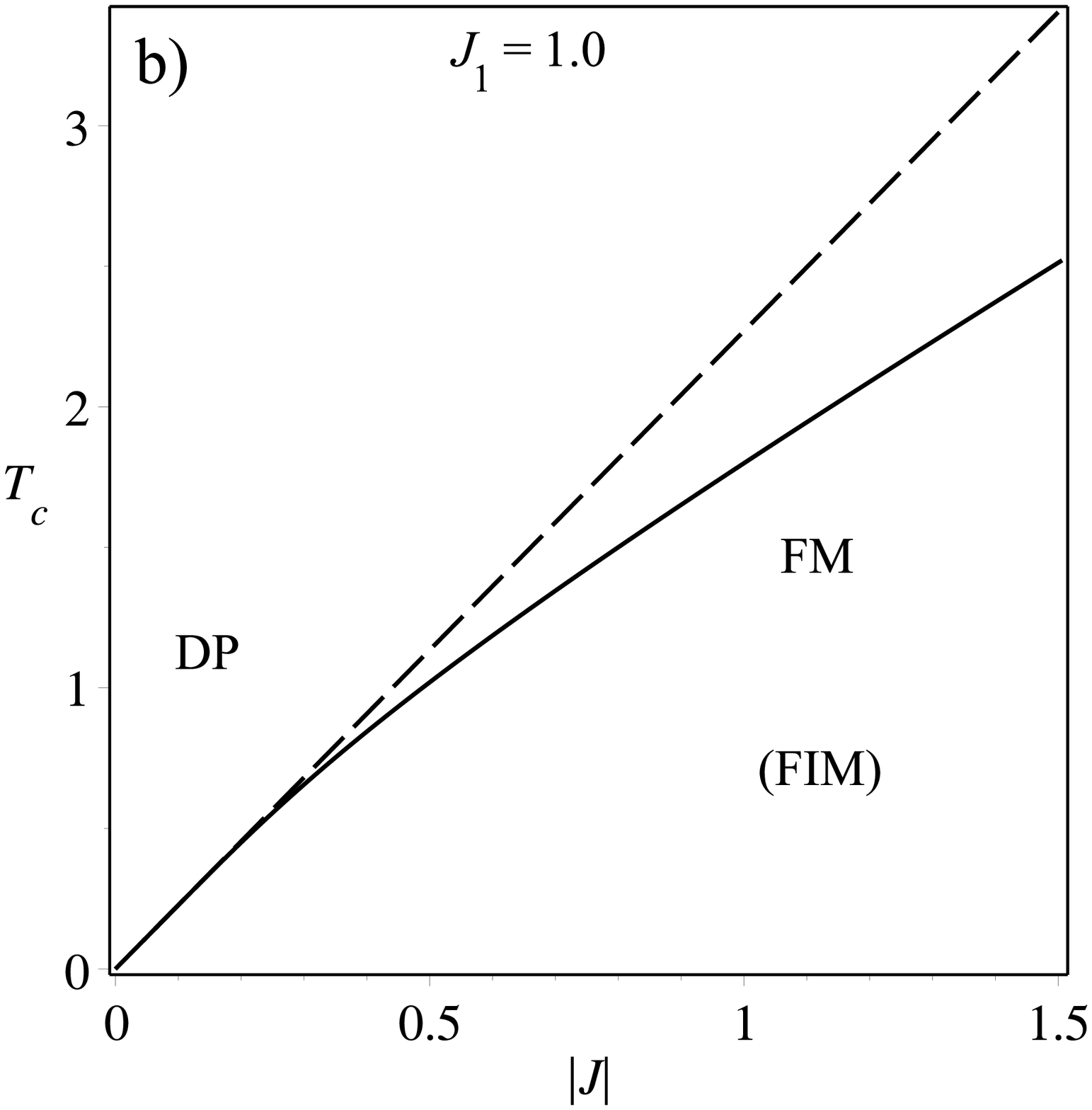} 
\par\end{centering}

\caption{\label{fig:phaseT} Phase diagrams as a function of temperature $T_{c}$
and the parameter $|J|$ for two different values of  $J_{1}$.
In (b) the dashed lines show the limiting values of $J\rightarrow0$
and $T_{c}\rightarrow0$.}
\end{figure}

\begin{figure}[h]
\begin{centering}
\includegraphics[scale=0.38]{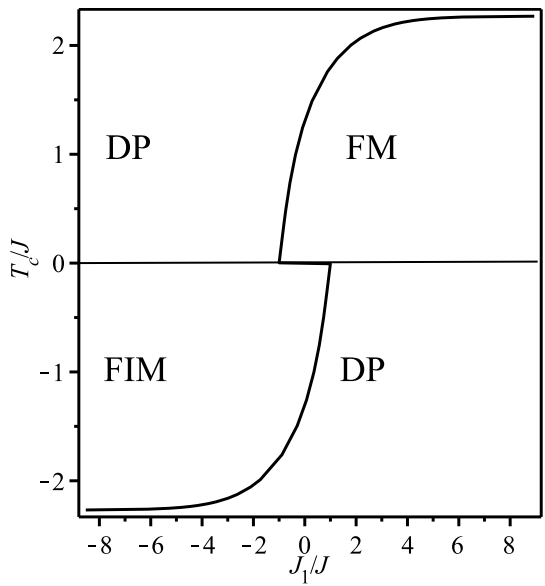} 
\par\end{centering}

\caption{\label{fig:grafico1} Phase diagram in the $(T_{c}/J,J_{1}/J)$ plane
for the pentagonal Ising model.}
\end{figure}

An alternative phase diagram is depicted in Fig. \ref{fig:phaseT},
where the phase diagrams are illustrated in the $(T_{c},|J|)$ plane
(since $T_{c}$ is invariant unde ther $J\rightarrow-J$ exchange) for
fixed parameter $J_{1}$. In Fig. \ref{fig:phaseT}(a) the second-order
phase transition line is shown in the $(T_{c},|J|)$ plane when the
parameter $J_{1}$  is fixed at $J_{1}=-1.0$;  in this case there
is a DP. Concretely, for $J<0$ we  show two regions:
 the FIM  phase and the   DP;  for $J>0$ we have a DP and a FM phase.

In Fig. \ref{fig:phaseT}(b) we show the behavior of the
critical temperature when $J_{1}=1.0$, displayed as a solid line.
For low values of $T\thickapprox0$,\textcolor{black}{{} the left-hand side of Eq. (\ref{eq:kk}) goes to infinity;  this implies that
the denominator on the right-hand side must satisfy the condition. $u_{c}^{12}-5u_{c}^{8}-5u_{c}^{4}+1=0$.
Thus we obtain the solution $T_{c}=\pm2.2691J$} (dashed lines), which
is valid within the limit $J\rightarrow0$ and $T_{c}\rightarrow0$.
This result is in agreement with the phase diagram
at zero temperature (see, for instance, Fig. \ref{fig:Ph-dgm}); more
specifically, for $J_{1}=1$, the phase transition occurs at $J=0$.

We now  comment on the finite-temperature phase diagrams
displayed in Fig. \ref{fig:grafico1}, in which the critical temperature
 $T_{c}/J$ is shown as a function of the parameter $J_{1}/J$.
Using the equation of the critical points (\ref{eq:kk}),\textcolor{magenta}{{}
}we obtain the plot illustrated in Fig. \ref{fig:grafico1}. From
that  we can analyze three limiting cases. 
\begin{description}
\item [{(i)}] For $\frac{J_{1}}{J}\rightarrow0$  we obtain $T_{c}/J=1.3084$
($T_{c}/J=-1.3084$) and  the pentagonal lattice is reduced to a ferromagnetic
(ferrimagnetic, with total magnetization $M=1/3$) decorated square
lattice. 
\item [{(ii)}] For $\frac{J_{1}}{J}\rightarrow\infty$, from our calculation,
for $J>0$ and $J_{1}>0$,  we find the solution $T_{c}/J=2.2691$;
in this case the pentagonal lattice is reduced to the ferromagnetic
square lattice. In the case $J<0$ and $J_{1}<0$ , in the limit under
consideration ($\frac{J_{1}}{J}\rightarrow\infty$), the pentagonal lattice falls into the
bottom righ- hand corner, which is a DP state (see Fig. \ref{fig:grafico1}). 
\item [{(iii)}] For $\frac{J_{1}}{J}\rightarrow-\infty$
we obtain $T_{c}/J=-2.2691$ for $J<0$ and $J_{1}>0$  and the pentagonal lattice  is
reduced to a ferrimagnetic (with total magnetization $M=1/3$) square
lattice. Meanwhile, for $\frac{J_{1}}{J}\rightarrow-\infty$ for
$J>0$ and $J_{1}<0$ the pentagonal lattice falls into a disordered state,
which corresponds to the top left  corner  Fig. \ref{fig:grafico1}. 
\end{description}
When $K_{2}=0$, according to a pentagonal lattice mapping onto
an effective square lattice (see Fig. \ref{fig:transf-pent}), within
the limit $\frac{J_{1}}{J}\rightarrow\infty$, the effective lattice
is reduced to a ferromagnetic square lattice. In the limit
$\frac{J_{1}}{J}\rightarrow-\infty$ the pentagonal Ising model reduces
to an antiferromagnetic square lattice.

\subsection{Internal Energy}

\begin{figure}
\includegraphics[scale=0.4]{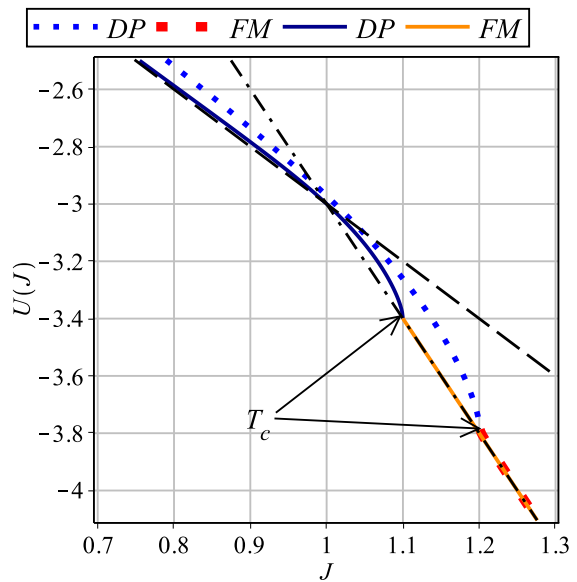}\caption{\label{intrn-energy}(Color online) Internal energy $U$ as a function of $J$ for a
fixed value $J_{1}=-1.0$. The dashed line corresponds to the DP ground energy and the dash-dotted line corresponds to the FM ground energy.}
\end{figure}

The internal energy of the pentagonal lattice Ising model defined by $U=T^{2}\frac{\partial\left(f/T\right)}{\partial T}$
can be obtained straightforwardly from Eq. \eqref{eq:mr}. In Fig.
\ref{intrn-energy}  we display the internal energy as a function
of coupling parameter $J$ in the low-temperature limit  in order
to observe the low-lying energy contribution close to the critical temperature
assuming $J_{1}=-1.0$. The dashed line corresponds to the DP internal
energy at $T=0$ (the ground-state energy), given by $U=-1-2J$, while
the dash-dotted line indicates the internal energy $U=1-4J$ from
which the phase transition at zero temperature occurs at $J=1$. In the case $T_{c}\rightarrow0$ and assuming $J=1+\delta$
with $\delta\gtrsim0$, we have $u_{c}\rightarrow\infty$ and then
it is possible to write $u_{c}=r_{c}^{-1-\delta}$. Further, by substituting
into Eq. (\ref{eq:kk}),  after some algebraic manipulations
we obtain $r_{c}^{2\delta}\approx\frac{1}{2\sqrt{2}}$, where $r_{c}=e^{-\frac{1}{T_{c}}}$
was defined previously. Thus a second-order phase transition occurs
at $T_{c}\thickapprox\frac{2(|J|-1)}{\ln(2\sqrt{2})}$. It is worth
 highlighting that the lowest critical temperature occurs at $T_{c}=0$.
Therefore, the contribution of the low-lying energy is absorbed by the
second-order phase transition. Certainly there is no second-order
phase transition for $J<1$ \textcolor{black}{and} $J_{1}=-1$. The
solid blue  (thick) line and the dark-blue  (thin) curve represent
the internal energy in the disordered phase, while by the red
(thick) line and the orange (thin) line curve represent the ferromagnetic
region for two critical temperatures $T_{c}=0.192$ and $0.385$.

\begin{figure}
\includegraphics[scale=0.42]{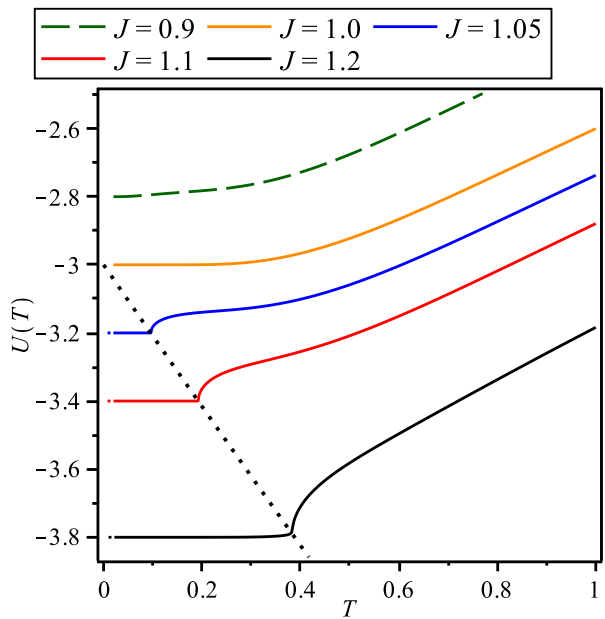}\caption{\label{fig:Int-enrg-UT}(Color online) Internal energy $U$ as a function of $T$
and a fixed value $J_{1}=-1.0$. The dotted  curve corresponds to $U(T_{c})$.}
\end{figure}

In Fig. \ref{fig:Int-enrg-UT} we plot the internal energy as a
function of temperature for several values of $J$ around the second-
order phase transition, assuming a fixed value for $J_{1}=-1.0$.
The black dotted line represents the internal energy $U(T_{c})$ evaluated
at the critical temperature $T_{c}$ given by Eq. \eqref{eq:lw-Tc} at
low temperature. Below the critical temperature, the internal energy is
almost constant (ferromagnetic phase), which means that it is mainly given
by the zero-temperature ground-state energy. Although at a critical
temperature there is a sudden change of curvature, this change becomes
dramatic for lower critical temperature;  for higher critical
temperature this change of curvature becomes smoother. The lowest
critical temperature occurs at $T_{c}=0$ for $J=1$; therefore, for
lower values of the coupling parameter $J$ there is no second-order
phase transition. For a sufficiently high temperature the internal energy
leads to an asymptotic limit, whereas the internal energy increases almost
proportionally to the temperature.

\subsection{The spontaneous magnetization}

The total magnetization $M$ is given by Eq. \eqref{eq:T-mag} for
the pentagonal Ising model, which will be discussed in order to show
the spontaneous magnetization. It ought to be pointed out that the
calculation of the magnetization of internal spin $s_{i}$, $i=1,2$,
may be obtained following the results obtained by Choy and Baxter
\cite{ch}, which could be expressed by the relation  
\begin{equation}
M_{1}=\langle s_{1}\rangle=c_{1}\langle\tau\rangle+c_{2}\langle\tau_{1}\tau_{2}\tau_{3}\rangle,
\end{equation}
where after some algebraic manipulations the coefficients become 
\begin{eqnarray}
c_{1} & = & \frac{1}{4}\left(\frac{\Sigma_{1}}{\xi_{1}}+2\frac{\Sigma_{5}}{\xi_{5}}\right),\\
c_{2} & = & \frac{1}{4}\left(\frac{\Sigma_{1}}{\xi_{1}}-2\frac{\Sigma_{5}}{\xi_{5}}\right).
\end{eqnarray}
Defining $\Sigma_{1}$ and $\Sigma_{5}$ in analogy to the Boltzmann
factors $\xi$, we have 
\begin{align}
\Sigma_{1}= & \frac{-2r}{u^{4}}+2ru^{4},\\
\Sigma_{5}= & \frac{-2r}{u^{2}}+2ru^{2}.
\end{align}

In order to obtain the three-spin correlation function $ $$\langle\tau_{1}\tau_{2}\tau_{3}\rangle$
we use a checkerboard Ising model  equivalent to that used by Choy and Baxter \cite{ch}.
Here we use the relation obtained in Ref. \cite{ch}, but in
our case we rewrite this relation in terms of the pentagonal Ising model
Boltzmann factor. Hence, using some algebraic manipulations we have
\begin{equation}
\langle\tau_{1}\tau_{2}\tau_{3}\rangle=R(r,u)\langle\tau_{1}\rangle,
\end{equation}
where 
\begin{equation}
R(r,u)=\tfrac{2\xi_{1}}{\xi_{1}-\xi_{3}}+\tfrac{\xi_{1}+\xi_{3}}{\xi_{1}-\xi_{3}}\left(1-\tfrac{2\xi_{1}\sqrt{\xi_{1}^{2}+\xi_{2}^{2}-2\xi_{2}\xi_{3}}}{\xi_{1}^{2}-\xi_{2}\xi_{3}}\right).
\end{equation}
Therefore, the magnetization $M_{1}$ can be expressed as a function
of magnetization $M_{0}$, which is given by 
\begin{equation}
M_{1}=\frac{1}{4}\left[\left(\frac{\Sigma_{1}}{\xi_{1}}+2\frac{\Sigma_{5}}{\xi_{5}}\right)+\left(\frac{\Sigma_{1}}{\xi_{1}}-2\frac{\Sigma_{5}}{\xi_{5}}\right)R(r,u)\right]M_{0}.\label{eq:mag3}
\end{equation}
It is important to highlight that Eq. \eqref{eq:mag3} is expressed
in terms of the original parameters of the pentagonal Ising model
instead of parameters of the effective Hamiltonian such as those obtained
by Urumov \cite{uru}. Using Eq. \eqref{eq:mag3}, we are easily
able to manipulate the parameters of the pentagonal Ising model in
order to discuss the spontaneous magnetization. At the critical point
we need to substitute the  expression $r=r_{c}^{\frac{T_{c}}{T}}$,
where $r_{c}$ is defined in Eq. \eqref{eq:kk}. From Eqs. \eqref{eq:mag2} and \eqref{eq:mag3} we can obtain a closed
expression for the total magnetization of the Ising model on a pentagonal
lattice, using the relation given by Eq. \eqref{eq:T-mag}.

\begin{figure}[h]
\begin{centering}
\includegraphics[scale=0.35]{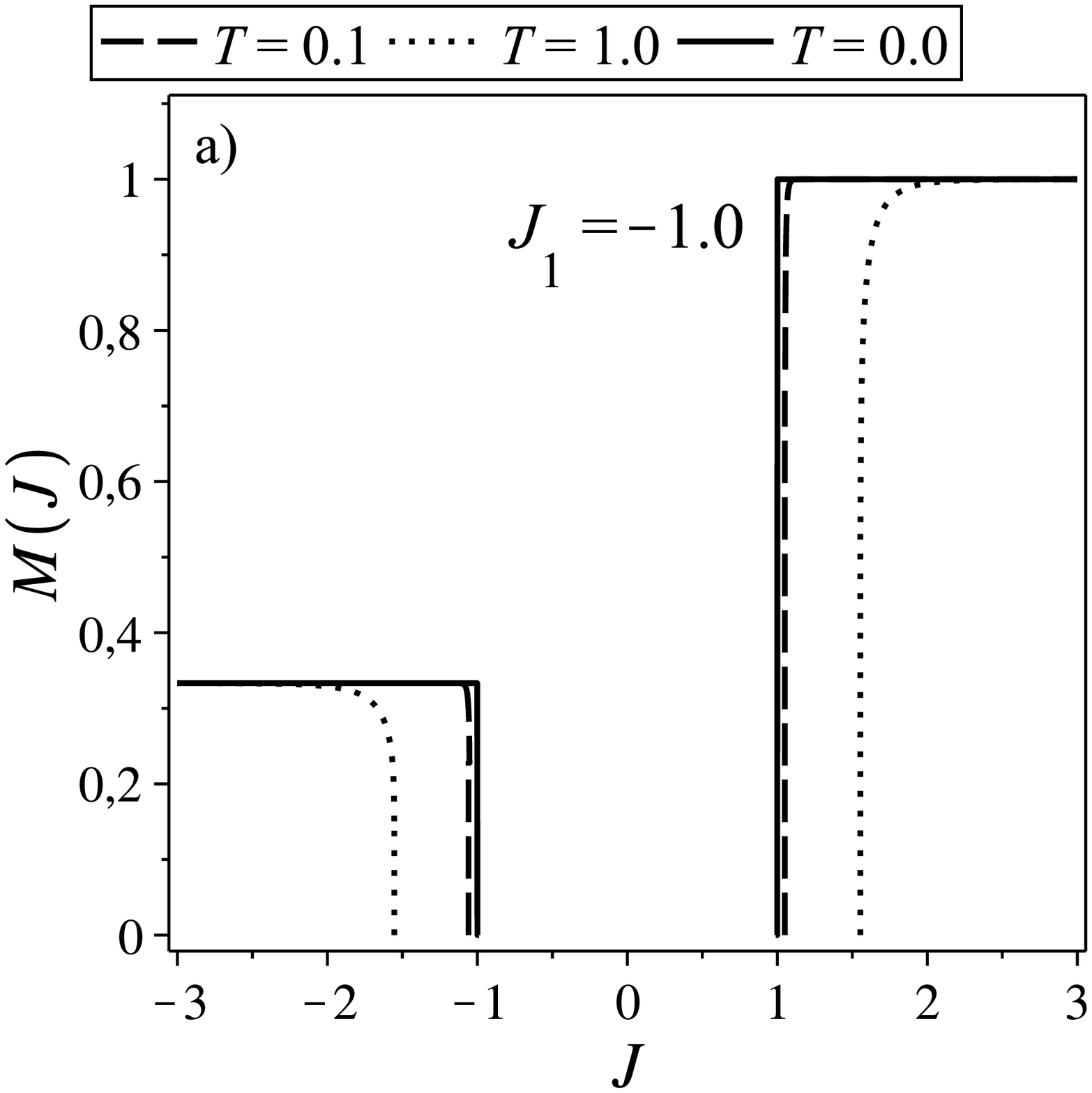} \includegraphics[scale=0.35]{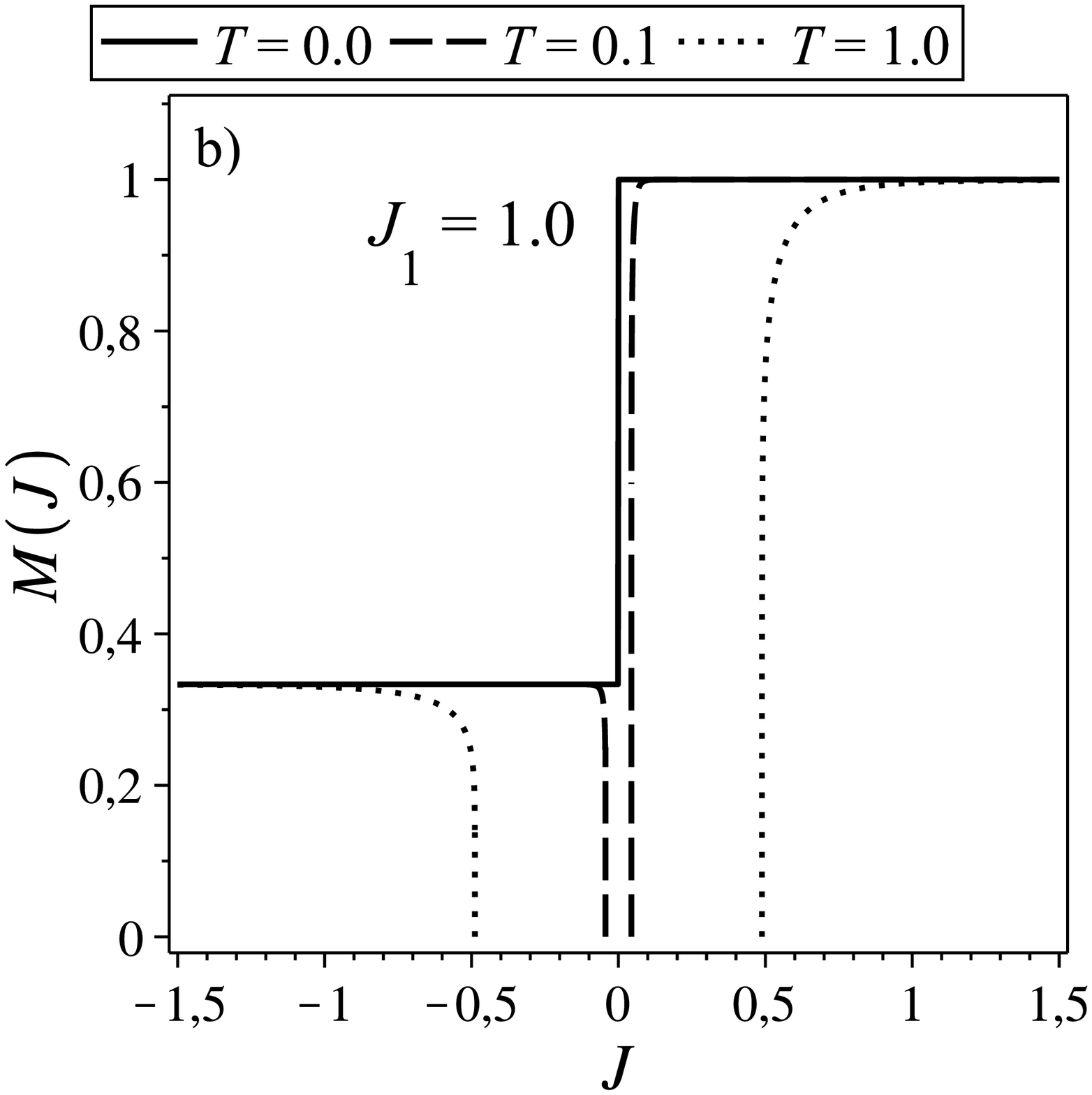} 
\par\end{centering}

\caption{\label{fig:T vs J} Total magnetization of the pentagonal Ising model
for three different values of $T$ as a function of the parameter $J$
and fixed parameter $J_{1}$. (a) $J_{1}=-1.0$ and (b) $J_{1}=1.0$.}
\end{figure}

We now discuss the behavior of the total magnetization of the pentagonal
Ising model as a function of the parameter $J$ for the low-temperature
limit. In Fig. \ref{fig:T vs J} we plot the magnetization at low
temperature as a function of $J$, where we display two types of 
plateaus for the FIM state and the FM state. This is in agreement with
the phase diagram displayed in Fig. \ref{fig:Ph-dgm} and \ref{fig:crit-points},
whereas the intermediate state corresponds to the FRU phase and the DP,
respectively.

Hence, in Fig. \ref{fig:T vs J}(a), for $J_{1}=-1.0$, when $T=0$
there are three well defined regions: FM phase with $M=1$, the FIM
phase with $M=1/3$, and the intermediate  FRU phase; also,
by increasing the temperature (for example, from $T=0.1$ to $1.0$)\textcolor{magenta}{{}
}the disordered phase increases\textcolor{green}{{} }(the $|J|$
increases). Meanwhile, in Fig. \ref{fig:T vs J}(b), for $J_{1}=1.0$
and at zero temperature, we have a direct phase transition between the
FM state and the FIM state. However, for a nearly zero temperature $T=0.1$,
a small intermediate region arises that  corresponds to the disordered
phase region. For higher temperature such as $T=1.0$, the DP region is even larger.

\begin{figure}
\begin{centering}
\includegraphics[scale=0.35]{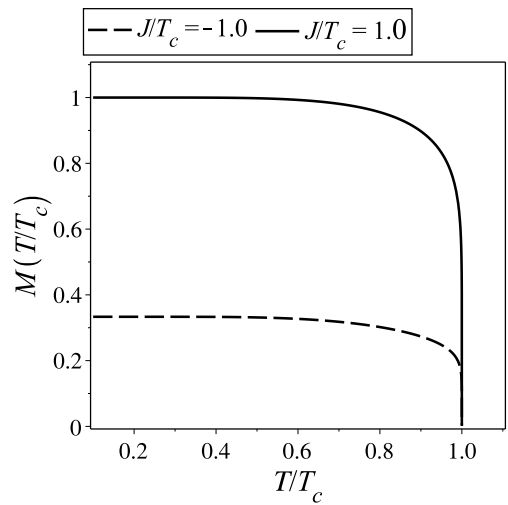} 
\par\end{centering}

\caption{\label{fig:M vc T} Temperature dependence of the total magnetization
of the pentagonal Ising model for two different values of $J/T_{c}$
and a fixed value $J_{1}=-1.0$.}
\end{figure}

Another way to analyze the total magnetization is by exploring
the temperature dependence $T/T_{c}$ of the total magnetization.
In Fig. \ref{fig:M vc T} we plot the magnetization as a function
of temperature for two values of $J/T_{c}=\pm1$ and $J_{1}=-1.0$,
where the magnetization of the saturated FM state and the FIM state is illustrated.
For the case of $J/T_{c}=1$ we have the FM region with $M=1$ 
the total magnetization at zero temperature. The total magnetization vanishes at $T=T_{c}$
as temperature increases; therefore, a second-order phase transition
occurs at the critical temperature.

For $J/T_{c}=-1$, the total magnetization corresponds to the FIM region, with $M=1/3$
at zero temperature, and it vanishes at $T=T_{c}$ as displayed in
Fig. \ref{fig:M vc T}. Thus a second-order phase transition occurs
again at the critical temperature, which is in agreement with the critical
point curve displayed in Fig. \ref{fig:crit-points}.

\subsection{Entropy }

The entropy can be easily obtained as a negative temperature derivative
from the free energy \eqref{eq:mr} $\mathcal{S}=-\frac{\partial f}{\partial T}$
, while the specific heat can be written as a temperature derivative
from the entropy $C=T\frac{\partial\mathcal{S}}{\partial T}$. In
what follows we  consider only the case $J>0$  since  for $J<0$
we have the same behavior.

\begin{figure}
\begin{centering}
\includegraphics[scale=0.28]{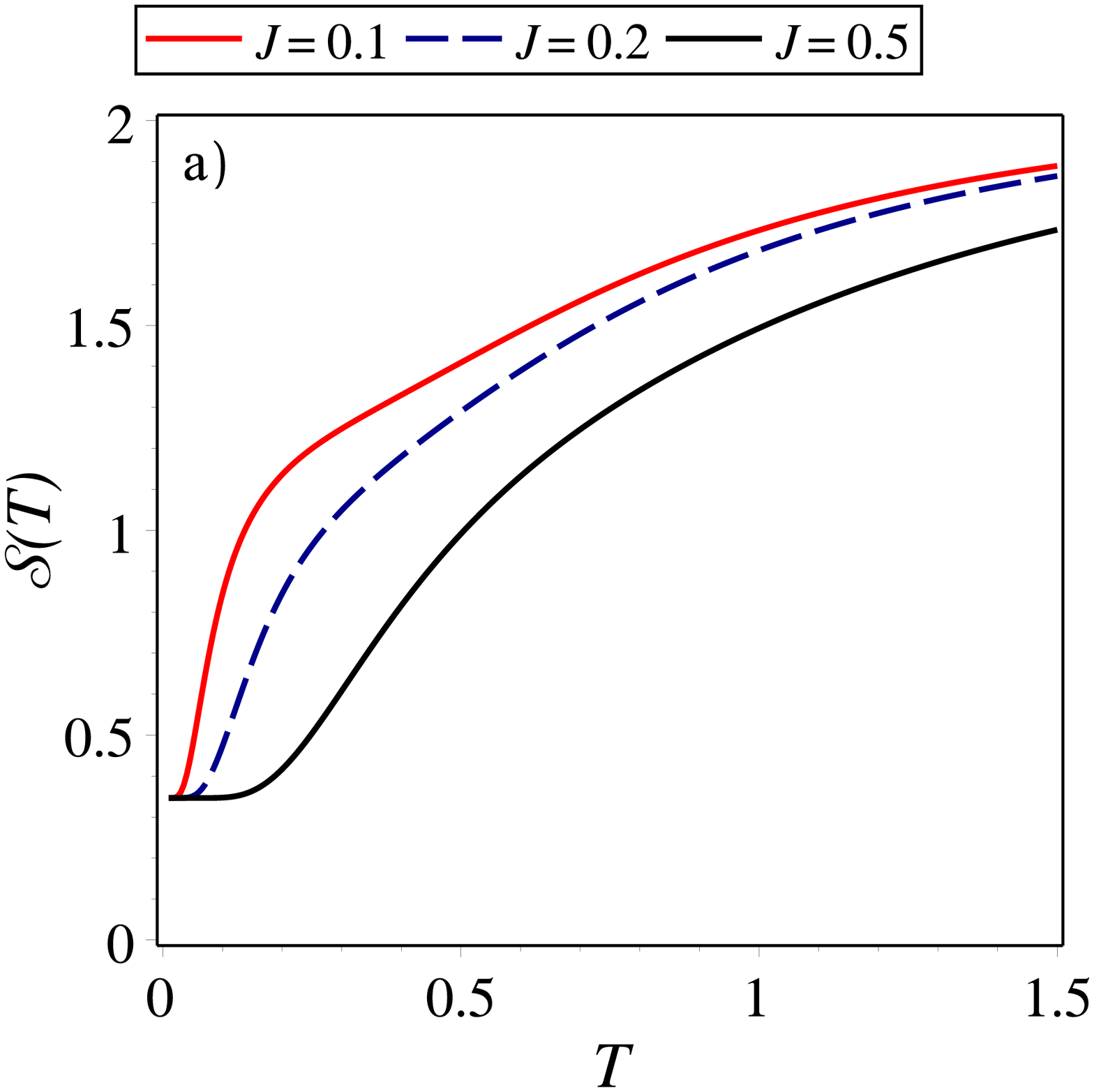} 
\par\end{centering}

\begin{centering}
\includegraphics[scale=0.28]{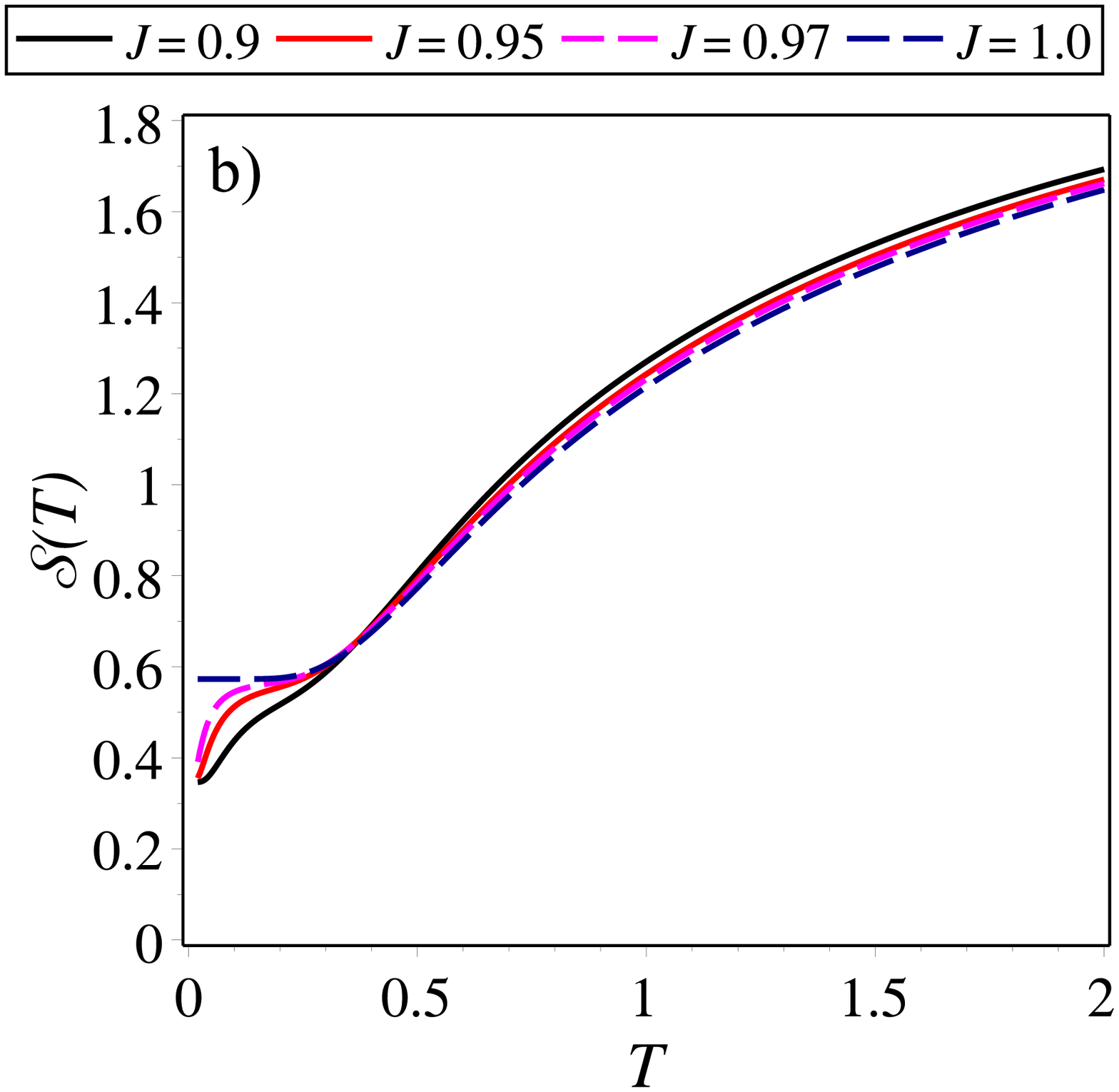} 
\par\end{centering}

\begin{centering}
\includegraphics[scale=0.28]{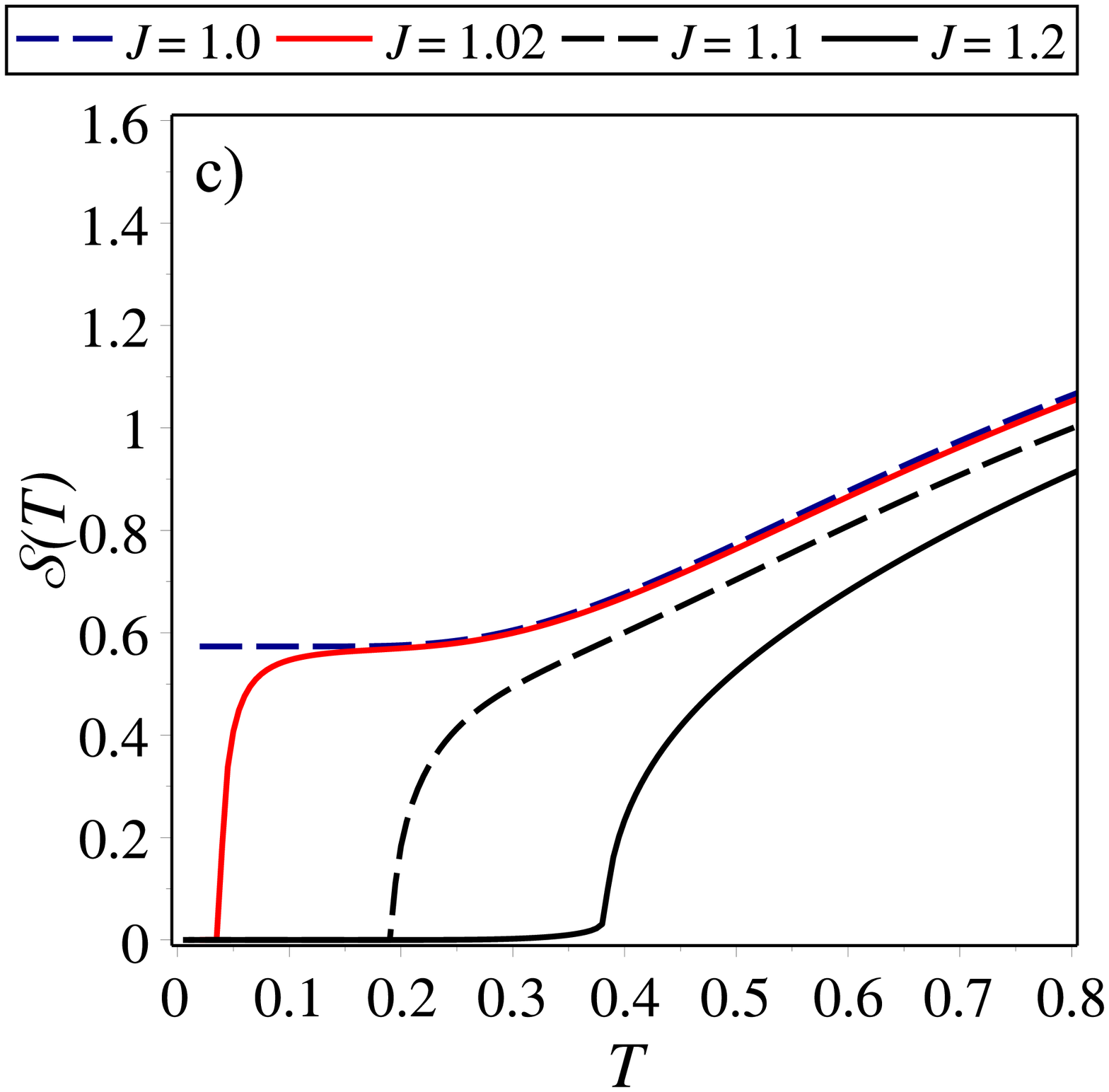} 
\par\end{centering}

\begin{centering}
\includegraphics[scale=0.28]{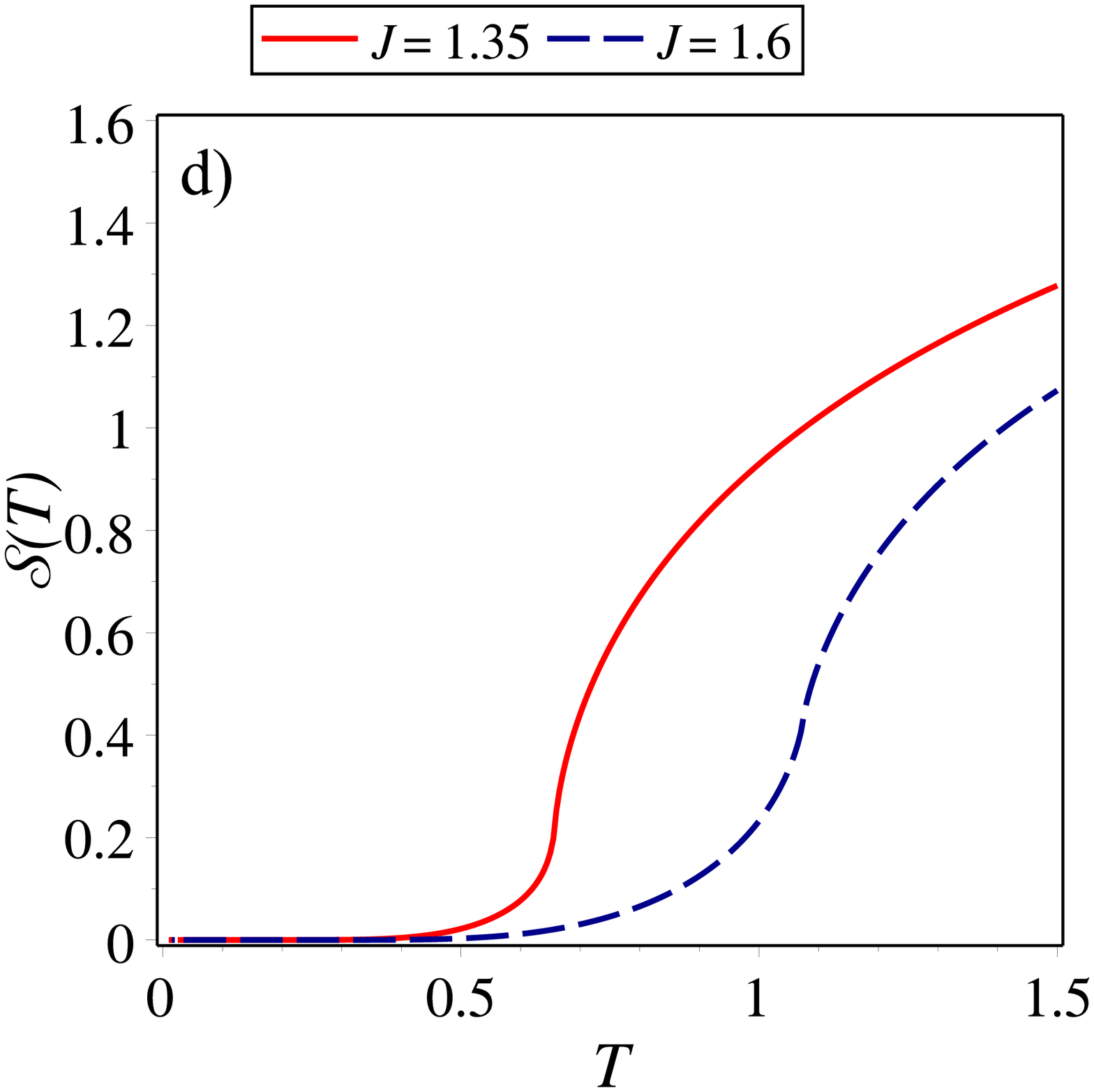} 
\par\end{centering}

\caption{\label{fig:Entro}(Color online) Entropy as a function of temperature for $J_{1}=-1.0$ and
(a) and (b) $J\leqslant1$ and (c) and (d) $J>1$}.
\end{figure}

In Fig. \ref{fig:Entro} we display the entropy as a function of
temperature for several values of $J$ with  $J_{1}=-1.0$ fixed. Figure
\ref{fig:Entro} (a) shows the low-temperature behavior of entropy, where the residual entropy appears at $\mathcal{S}_{0}=\ln(2)/2=0.3465$
for $|J|<1.0$. This means that we are in a geometrically frustrated
region, which is in agreement with the illustration of the phase diagram
in Fig. \ref{fig:Ph-dgm}. The residual entropy is proportional
to $\ln\left(2\right)$; this number comes from the two configurations
given in Eq.\eqref{eq:fru-st}. For $|J|=1.0$ the residual entropy has a different nontrivial value as displayed
in Fig. \ref{fig:Entro}(b).\textcolor{black}{{} To obtain the
entropy explicitly, we return to Eqs. (\ref{eq:a1}) and (\ref{eq:q1}) and
set $r_{c}=\frac{1}{u_{c}}=e^{-1/_{T_{c}}}$. Thus  we obtain 
\begin{eqnarray*}
r_{c}^{6}A(\phi) & = & \frac{5}{2}+\sqrt{2}\cos(\phi)+O(r_{c}^{2})
\end{eqnarray*}
and 
\begin{eqnarray*}
r_{c}^{12}Q & = & \frac{1}{4}+\sqrt{2}\cos(\phi)+2\cos^{2}(\phi)+O(r_{c}^{2}).
\end{eqnarray*}
Finally, for $T_{c}\rightarrow0$ ($r_{c}\rightarrow0$) and using
Eq. (\ref{eq:mr}), the residual entropy becomes 
\begin{eqnarray}
\mathcal{S}_{0} & = & \frac{1}{4\pi}\int_{0}^{2\pi}\ln(\tfrac{5}{2}+|\sqrt{2}\cos(\phi)+\tfrac{1}{2}|+\sqrt{2}\cos(\phi))\mathrm{d}\phi\nonumber \\
 & \approx & 0.5732714757.
\end{eqnarray}
} This is due to the degeneration of the phase boundary between the FRU
and FM (or FIM) regions at $T=0$ (for detail see Fig. \ref{fig:Ph-dgm}).
This result was derived in a  way similar to that  discussed by Wannier \cite{wannier}
for the case of a two-dimensional triangular lattice. While in Fig.
\ref{fig:Entro}(c) there is no residual entropy for $|J|>1$, the
standard temperature dependence of entropy appears with a strong change
of curvature located at critical points where  a second-order
phase transition occurs.

It ought to be highlighted that in the low-temperature limit (below the
critical temperature) the entropy (in the FM state) for $J>1$ and with
$J_{1}=-1$ can be obtained from Eq. \eqref{eq:mr}.
More explicitly, by fixing $J=1+\delta$, where $\delta>0$, it is possible
to write $u=\frac{1}{rs}$, with $r={\rm e}^{\frac{-1}{T}}$ and $s={\rm e}^{\frac{-\delta}{T}}$.
Substituting Eqs. (\ref{eq:a1}) and (\ref{eq:q1}) into   Eq. (\ref{eq:mr}),
the integration of Eq. (\ref{eq:mr})  results in

\begin{eqnarray*}
f & \approx & -3-2\delta-T(-2\ln s+2r^{2}s^{4})\\
 & \approx & 1-4J-2T{\rm e}^{\frac{-2(2|J|-1)}{T}}.
\end{eqnarray*}
Finally, the entropy $ $$\mathcal{S}=-\frac{\partial f}{\partial T}$
can be written as

\begin{equation}
\mathcal{S}\thickapprox\frac{2\Delta}{T}\mathrm{e}^{-\Delta/T},
\end{equation}
 where $\Delta\equiv2(2|J|-1)$ is the energy gap. In Figs. \ref{fig:Entro}(c) and  \ref{fig:Entro}( d) the low-temperature limit is well
fitted by the above limiting expression.

\begin{figure}
\includegraphics[scale=0.42]{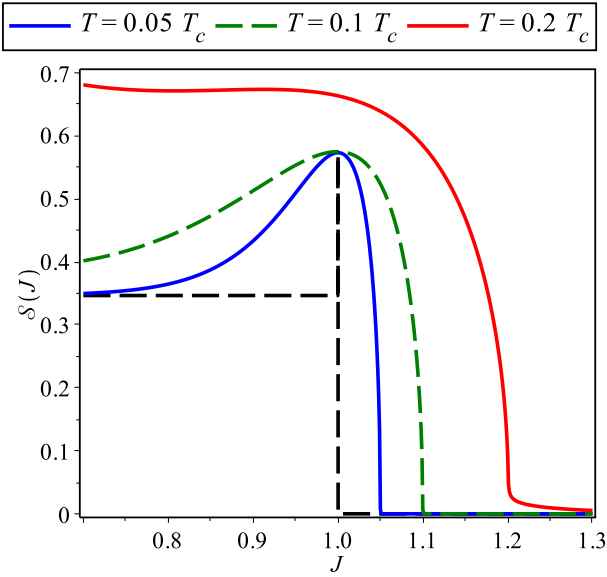}\caption{\label{fig:S-J-fxd-T}(Color online)  Entropy $\mathcal{S}$ as a function of the
parameter $J$  for a fixed low temperature and $J_{1}=-1.0$.}
\end{figure}

An additional plot of entropy $\mathcal{S}$ against $J$ in the low-temperature limit is displayed
in Fig. \ref{fig:S-J-fxd-T},  where the residual entropy is illustrated by the dashed black line at zero
temperature. For $J<1$ there is residual entropy $\mathcal{S}=\ln(2)/2$,
while for $J=1$ the residual entropy becomes $\mathcal{S}=0.5732714757$;
for higher values of $J>1$ there is no residual entropy. Thereafter,
we observed the entropy in the low-temperature limit, where we can
show the effects of residual entropy. The low-lying energy contribution
for the entropy between the DP and the FM phase is absorbed by the second-order phase
transition as a consequence of  the entropy falling dramatically
to zero entropy for $J>1$; for higher temperature the entropy change
curvature becomes softer.

\subsection{Specific heat}

\begin{figure}
\begin{centering}
\includegraphics[scale=0.3]{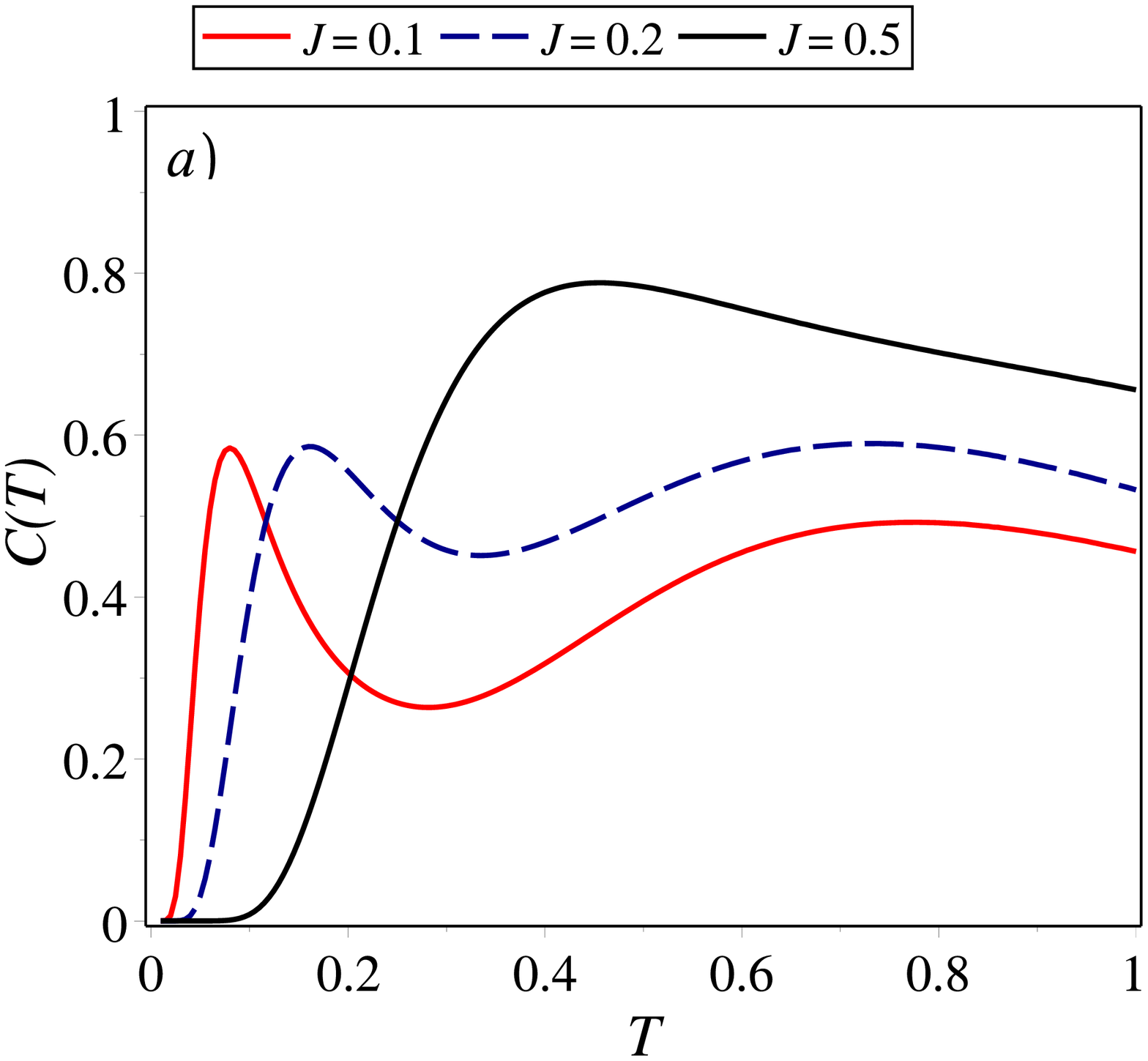} 
\par\end{centering}

\begin{centering}
\includegraphics[scale=0.3]{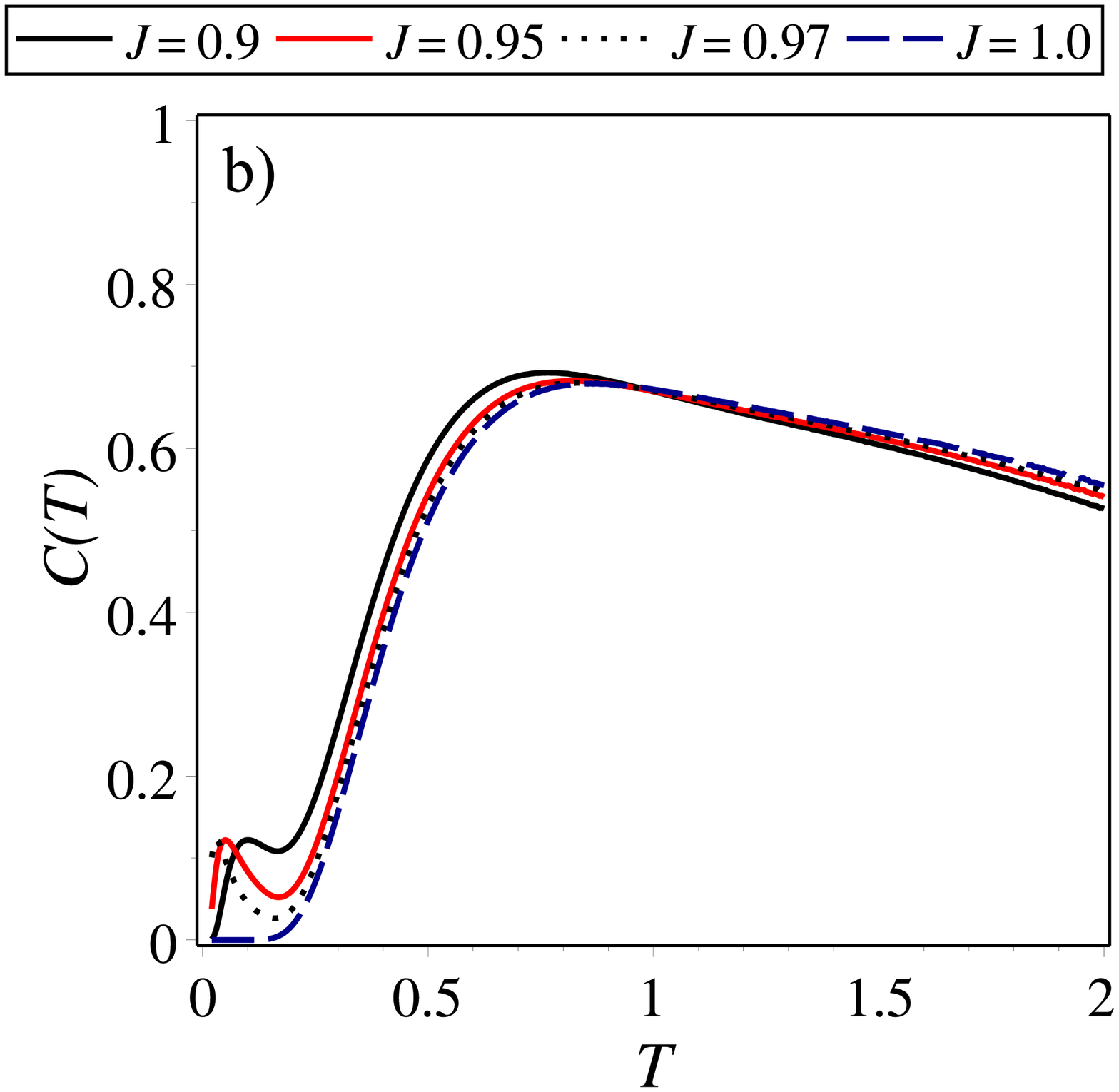} 
\par\end{centering}

\begin{centering}
\includegraphics[scale=0.3]{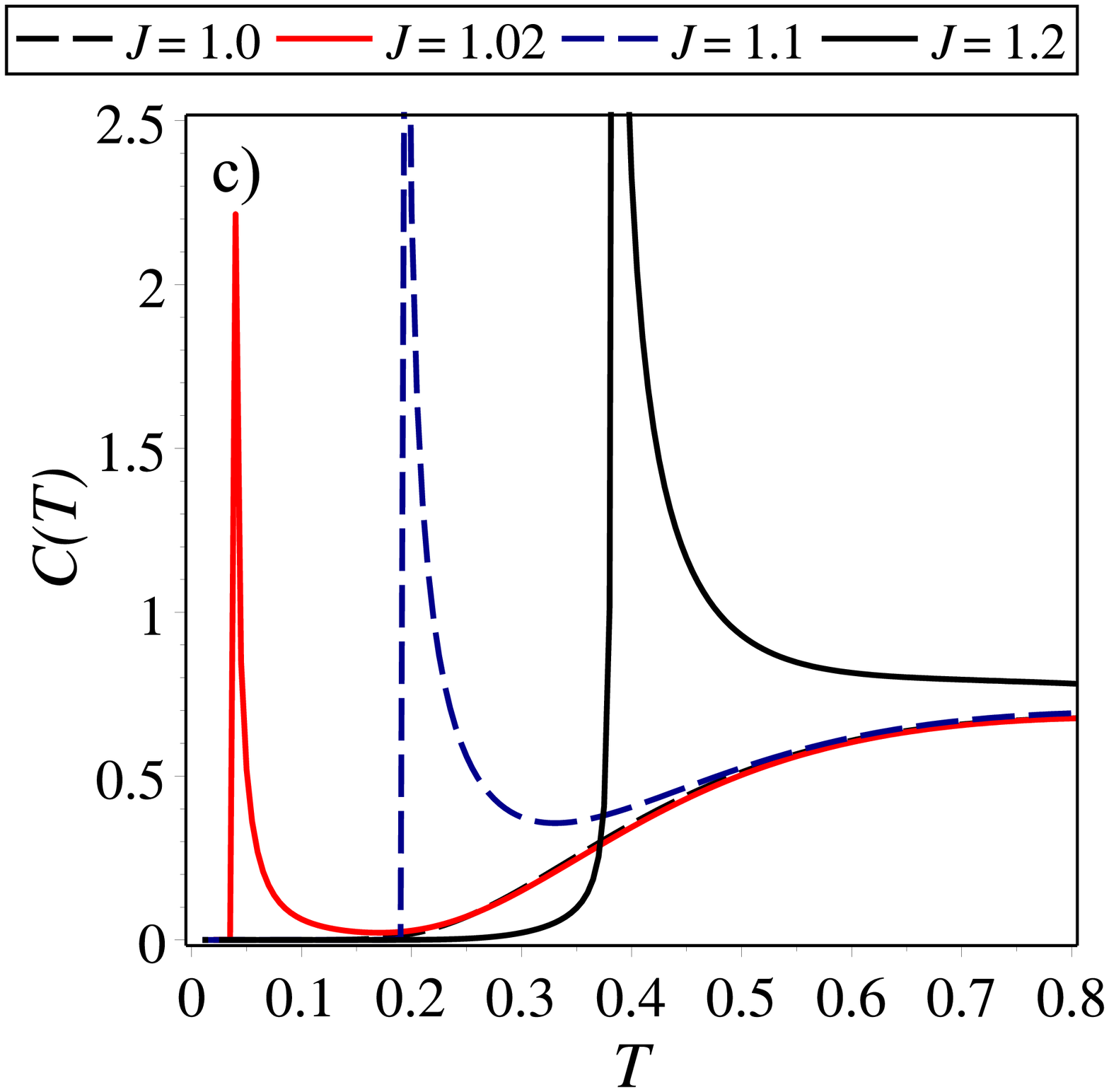} 
\par\end{centering}

\begin{centering}
\includegraphics[scale=0.3]{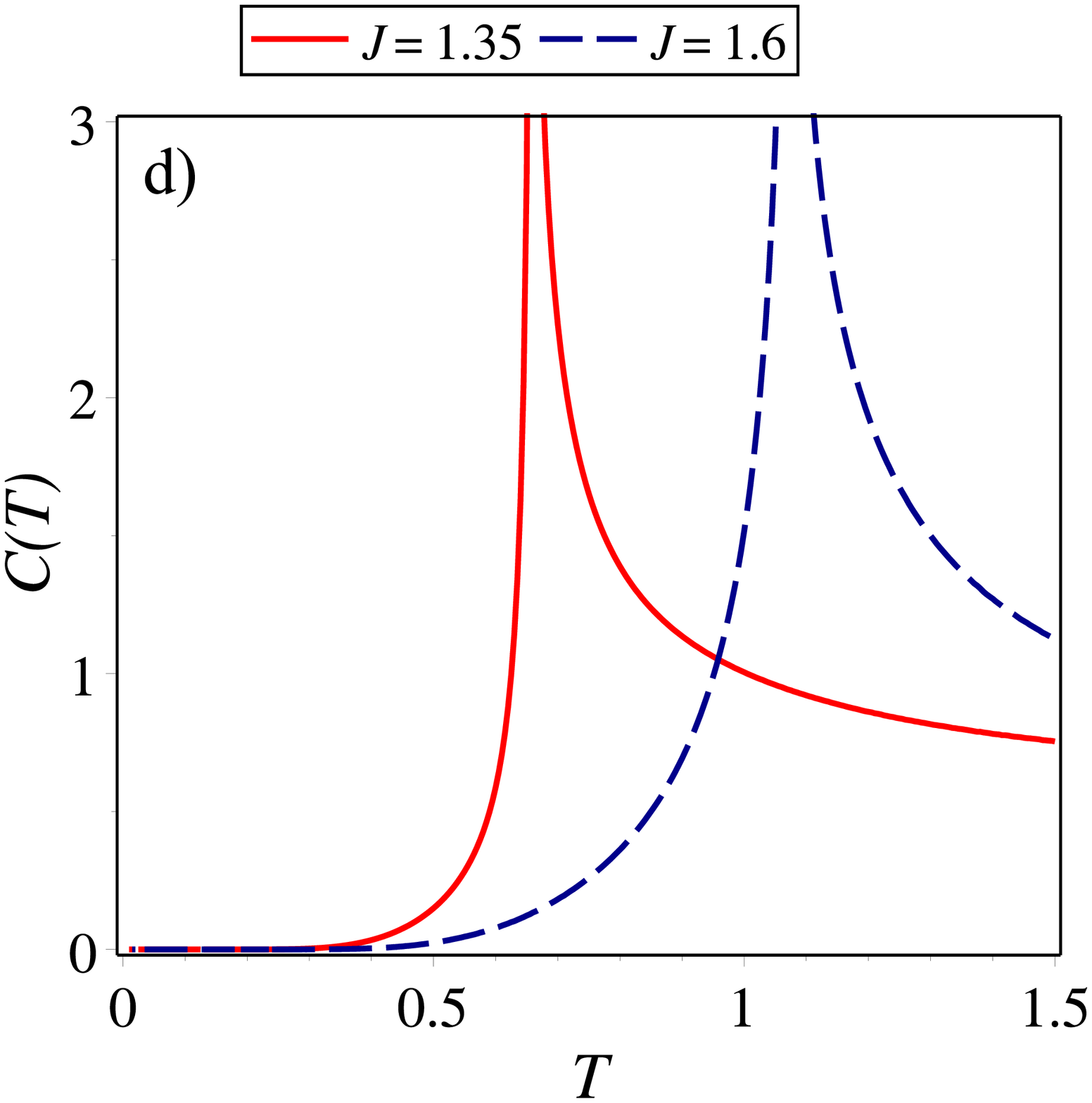} 
\par\end{centering}

\caption{\label{fig:Calor} (Color online) Specific heat as a function of temperature
for $J_{1}=-1.0$ and (a) and (b) $J\leqslant1$ and (c) and (d) $J>1$.}
\end{figure}

Finally, we conclude our analysis of thermodynamics by exploring the temperature
dependence of the specific heat. Some typical
thermal variations of the specific heat of the pentagonal Ising model
are plotted in Fig. \ref{fig:Calor}  for several values of $J$
and  $J_{1}=-1.0$ fixed. In Figs. \ref{fig:Calor}(a) and (b) we present
the temperature dependence of the specific heat in the DP states
and  show that there is no phase transition at finite temperature  because
we are observing the frustrated region. In Figs. \ref{fig:Calor}(c) and (d)
the specific heats are logarithmically divergent at the critical temperature,
which is associated with a continuous phase transition between the spontaneously
ordered and disordered phases. Clearly, this means that we are facing a
FM region, which can be verified in the phase diagram illustrated in
Fig. \ref{fig:Ph-dgm}.

Plotted in Fig. \ref{fig:Calor}(c) when $J\gtrsim1$ and $J_{1}=-1$
is  the specific heat versus temperature; a surprisingly almost null specific heat is displayed, until the critical
temperature is achieved. When the absolute value of the exchange interaction
is only slightly above 1, the critical temperature (an order-disorder
phase transition) can be obtained easily from Eq. \eqref{eq:kk}; this
critical temperature is approximately given by the expression

\begin{equation}
T_{c}\thickapprox\frac{2(|J|-1)}{\ln(2\sqrt{2})}\label{eq:lw-Tc}
\end{equation}
in the low-temperature limit.

\begin{figure}
\includegraphics[scale=0.4]{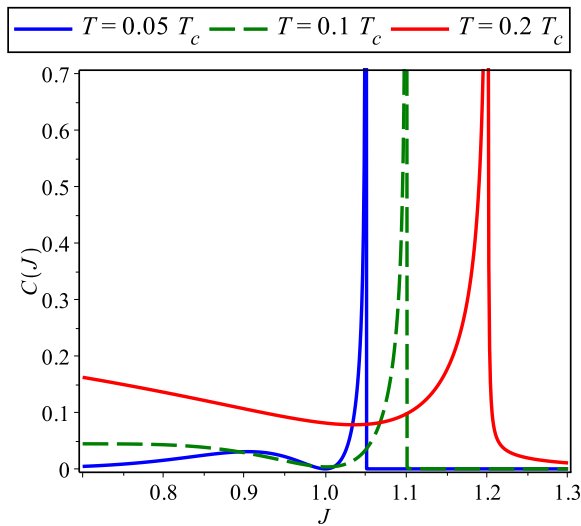}\caption{\label{fig:CvsJ-fxd-T}(Color online) Specific heat as a function of $J$  in
the low-temperature limit and $J_{1}=-1.0$.}
\end{figure}

In Fig. \ref{fig:CvsJ-fxd-T} we display the specific heat as a
function of $J$ in the low-temperature limit, where we show the
specific heat behavior around the second-order phase transition. As
 discussed previously, the low-lying energy contribution is absorbed
by a second-order phase transition in the case of $J>1$, while for $J<1$
there is no second-order phase transition; then we can observe the
low-lying energy contribution as a small anomalous broad peak.

Similarly, the low-temperature asymptotic limit for specific heat can
be derived from Eq. \eqref{eq:mr} so that

\begin{equation}
C\thickapprox\frac{2\Delta^{2}}{T^{2}}\mathrm{e}^{-\Delta/T}.\label{eq:Lw-spc-heat}
\end{equation}

The energy gap is large enough even when $J=1.02$ because the order-disorder
transition occurs for $|J|\thickapprox1+\frac{T_{c}}{2}\ln(2\sqrt{2})$.

\begin{figure}
\includegraphics[scale=0.4]{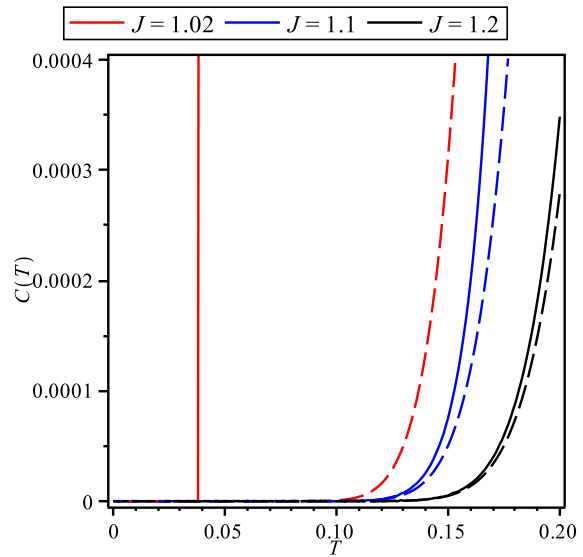}\caption{\label{fig:Lw-spc-ht}(Color online) Low-temperature limit specific heat against
temperature for the same values as in Fig. \ref{fig:Calor}(c). The solid
line corresponds to the exact specific heat and the dashed line is the low-temperature limit of the specific heat.}
\end{figure}

In Fig. \ref{fig:Lw-spc-ht} we show the magnification of Fig.
\ref{fig:Calor}(c) in low-temperature limit, which is well fitted
by Eq.\eqref{eq:Lw-spc-heat}. The solid line corresponds to the exact
specific heat and the dashed line  represents the low-temperature
asymptotic limit of the specific heat. For $J=1.02$ the corresponding
low-temperature approximation is valid for $T<0.0385$, while for
$J=1.1$ and $J=1.2$ clearly the low-temperature curve accompanies
quite well the exact solution.

\section{Conclusions}

Using the direct decoration transformation \cite{ono}, we have solved
the pentagonal Ising model with a more general coupling parameter
and compared it with Urumov's \cite{uru} solutions. We have found a
frustrated phase of the pentagonal Ising model.

In addition, we have obtained a simplified solution for the free energy,
as well as a closed expression for the critical temperature. Although
this model has already been solved by Urumov through the standard
decoration transformation \cite{fisher,phys-A-09} in the nonfrustrated
region ($J_{1}>0$), such a result contains unnecessary intermediate
parameters that can be avoided so that a closed expression similar
to Eq. \eqref{eq:kk} can be obtained.  We have studied the ground-state phase diagram, which exhibits a ferromagnetic state, a ferrimagnetic
state, and a frustrated state at $J_{1}<-|J|$. Following the
exact solution for the pentagonal Ising model, we have discussed the
finite-temperature phase diagram, as shown in Figs. \ref{fig:phaseT}
and \ref{fig:grafico1}, identifying the phase transition between
the FIM state and the DP state and also between the DP state and
the FM state.

The analysis of the limits in Fig. \ref{fig:grafico1} allows one
to find three relevant phases. For $J_{1}\rightarrow0$ the pentagonal
lattice reduces to a ferromagnetic (ferrimagnetic) decorated square
lattice. For $J_{1}\rightarrow\infty$ and $J>0$ the pentagonal
lattice reduces to the ferromagnetic square lattice. Finally, for
$J_{1}\rightarrow-\infty$ and $J<0$ the pentagonal Ising model
reduces to a ferrimagnetic square lattice. The total magnetization
as a function of the parameter $J$ and for a fixed value of $J_{1}$
for the ferromagnetic state $(M=1)$ and the ferrimagnetic state $(M=1/3)$
is shown in Fig. \ref{fig:T vs J}.

For a fixed value of $J_{1}=-1$ there is a residual entropy $\mathcal{S}_{0}=0.3465.$
For $|J|<1.0$ and  $|J|=1.0$  a nontrivial residual entropy
$\mathcal{S}_{0}=0.5732$ is found, as shown in Fig. \ref{fig:Entro}.
Because of the frustrated state, the entropy below the critical temperature
shows a strong change of curvature for $J\gtrsim1$. The specific
heat capacity was also investigated at fixed $J_{1}=-1$ and $|J|<1$
[see Figs. \ref{fig:Calor}(a) and \ref{fig:Calor}(b)]. For $J_{1}=-1$ and $|J|\gtrsim1$
we have unusual behavior due to frustration of the entropy and the
heat capacity at temperatures below the critical value, as shown in
Figs. \ref{fig:Calor}(c) and \ref{fig:Calor}(d).

\section{Acknowledgment}

M.R. acknowledges FAPEMIG for financial support. O.R. and
S.M.d.S. thank CNPq and FAPEMIG for partial financial support.


\begin{thebibliography}{10}
\bibitem{on} L. Onsager, Phys. Rev. \textbf{65} 117  (1944).

\bibitem{gr} H. S Green and C. A. Hurst, 1964 Order-Disorder Phenomena
( Interscience, New York, 1964).

\bibitem{ho} R. M. F. Houtappel, Physica. \textbf{16} 425 (1950);
K. Husimi and I. Syozi, Prog. Theor. Phys. \textbf{5} 177  (1950).

\bibitem{sy} I. Syozi, Prog. Theor. Phys. \textbf{6} 306 (1951).

\bibitem{uti} T. Utiyama, Prog. Theor. Phys. \textbf{6} 907 (1951).

\bibitem{va} V. G. Vaks, A. I. Larkin, and N. Yu. Ovchinnikov, Zh.
Eksp. Teor. Fiz. \textbf{49} 1180 (1965).

\bibitem{sun} F. Sun, X. M. Kong, and X. Ch. Yin, Commun. Theor. Phys.
\textbf{45} 555 (2006).

\bibitem{wal} M. H. Waldor, W. F. Wolff, and J. Zittartz, Z. Phys.
B Condensed. Matter. \textbf{59} 43 (1985).

\bibitem{uru} V. Urumov, J. Phys. A: Math. Gen. \textbf{35} 7317 (2002).

\bibitem{fisher}M. E. Fisher, Phys. Rev. \textbf{113} 969 (1959).

\bibitem{cairo} E. Ressouche, V. Simonet, B. Canals, M. Gospodinov,
and V. Skumryev, Phys. Rev. Lett. \textbf{103} 267204 (2009).

\bibitem{Iliev} M. N. Iliev, A. P. Litvinchuk, V. G. Hadjiev, M. M. Gospodinov, V. Skumryev, and E. Ressouche, Phys. Rev. B \textbf{81} 024302 (2010).

\bibitem{Liu}T. Liu, Y. Xu, C. Zeng, Mater. Sci. Eng. B \textbf{176} 535
(2011).

\bibitem{K Jin}K. Jin, B. Luo, S. Zhao, J. Wang, C. Chen, Chin. Phys.
Lett. \textbf{28} 087301 (2011).

\bibitem{cairo-hubb}A. Ralko,  Phys. Rev. B \textbf{84} 184434
(2011).

\bibitem{ono} O. Rojas, and S. M. de Souza, J. Phys. A: Math. Theor.
\textbf{44} 245001 (2011).

\bibitem{phys-A-09}O. Rojas, J. S. Valverde and S. M. de Souza, Physica
A \textbf{388} 1419 (2009).

\bibitem{strecka-pla}J. Strecka, Phys. Lett. A, \textbf{374} 3718 (2010).

\bibitem{strecka-mixd}J. Strecka, L. Canova and M. Jascur, Phys.
Rev. B \textbf{76} 014413 (2007); A. Dakhama, Physica A \textbf{252} 225
(1998).

\bibitem{loh}Y. L. Loh, D. X. Yao and E. W. Carlson, Phys. Rev. B
\textbf{77} 134402 (2008).

\bibitem{strecka-triang}J. Strecka, L. Canova, M. Jascur and M. Hagiwara,
Phys. Rev. B \textbf{78} 024427 (2008).

\bibitem{our-4-6-latt}J. S. Valverde, O. Rojas, and S. M. de Souza,
Phys. Rev. E \textbf{79} 041101 (2009).

\bibitem{Fan-wu}C. Fan and F. Y. Wu, Phys. Rev. B \textbf{2} 723 (1970).

\bibitem{ch} T. C. Choy and R. J. Baxter, Phys. Lett. A. \textbf{125} 365
(1987).

\bibitem{baxter-free-ferm}R. J. Baxter, Proc. R. Soc. London Ser. A \textbf{404} 1
(1986).

\bibitem{wannier}G. H. Wannier, Phys. Rev. \textbf{79} 357 (1950).\end{thebibliography}
\end{document}